\documentclass[%
 reprint,
 amsmath,amssymb,
 aps,
]{revtex4-2}

\usepackage{graphicx}
\usepackage{dcolumn}
\usepackage{bm}
\usepackage{xcolor}
\usepackage{placeins}
\usepackage{gensymb}
\usepackage[colorlinks=True, allcolors=blue]{hyperref}

\begin{document}

\preprint{APS/123-QED}

\title{Chiplet technology for large-scale trapped-ion quantum processors}

\author{Bassem Badawi$^\text{1}$}
\author{Philip C. Holz$^\text{2}$}
\author{Michael Raffetseder$^\text{1}$}
\author{Nicolas Jungwirth$^\text{1}$}
\author{Juris Ulmanis$^\text{2}$}
\author{Hans-Joachim Quenzer$^\text{3}$}
\author{Dirk Kähler$^\text{3}$}
\author{Thomas Monz$^\text{1,2}$}
\author{Philipp Schindler$^\text{1}$}
\affiliation{%
$^\text{1}$University of Innsbruck, Institute of Experimental Physics, Technikerstraße 25/4, 6020 Innsbruck, Austria\\
$^\text{2}$Alpine Quantum Technologies GmbH, Technikerstraße 17/1, 6020 Innsbruck, Austria\\
$^\text{3}$Fraunhofer Institute for Silicon Technology ISIT, Fraunhoferstraße 1, 25524 Itzehoe, Germany\\
}%

\date{\today}

\begin{abstract}
Trapped ions are among the most promising platforms for realizing a large-scale quantum information processor. Current progress focuses on integrating optical and electronic components into microfabricated ion traps to allow scaling to large numbers of ion qubits. Most available fabrication strategies for such integrated processors employ monolithic integration of all processor components and rely heavily on CMOS-compatible semiconductor fabrication technologies that are not optimized for the requirements of a trapped-ion quantum processor. In this work, we present a modular approach in which the processor modules, called chiplets, have specific functions and are fabricated separately. The individual chiplets are then combined using heterogeneous integration techniques. This strategy opens up the possibility of choosing the optimal materials and fabrication technology for each of the chiplets, with a minimum amount of fabrication limitations compared to the monolithic approach. Chiplet technology furthermore enables novel processor functionalities to be added in a cost-effective, modular fashion by adding or modifying only a subset of the chiplets. We describe the design concept of a chiplet-based trapped-ion quantum processor and demonstrate the technology with an example of an integrated individual-ion addressing system for a ten-ion crystal. The addressing system emphasizes the modularity of the chiplet approach, combining a surface ion trap manufactured on a glass substrate with a silicon substrate carrying integrated waveguides and a stack of 3D-printed micro-optics, achieving diffraction-limited focal spots at the ion positions.   
\end{abstract}

\maketitle


\section{Introduction}
\label{sec: intro}
Trapped ions are a promising candidate to realize a scalable quantum computing platform. Recent research in the field shows record qubit gate fidelity \cite{Hughes.2025, Smith.2025}, long coherence times \cite{Wang.11.01.2021} and the highest quantum volume achieved for any quantum computing platform so far \cite{Moses.2023}. Error-correcting methods have been applied to protect the quantum information from noise, demonstrating fault-tolerant universal gate sets and state teleportation on the logical qubit level with macroscopic and microfabricated ion trap architectures \cite{Postler.2022,Ryananderson.19.09.2024}. Numerous quantum computing applications have been implemented, such as quantum chemistry \cite{Ollitrault.2024.04.24}, quantum many-body dynamics \cite{Kazuhiro.2025}, combinatorial optimization \cite{Zhu.2022} or quantum machine learning \cite{Zhu.2023}. 

For quantum computers to become useful for such applications, it is generally believed that more than thousands of qubits are needed \cite{Proctor.2025}. State-of-the-art trapped-ion quantum computers (TIQCs) are inspired by the quantum charged-coupled device (QCCD) architecture \cite{Kielpinski.2002}, where the qubit register is distributed over multiple ion crystals, each acting as a qubit sub-register, and ion crystal reconfiguration techniques are used to achieve all-to-all qubit connectivity \cite{Blakestad.2009, Pino.2021, Tinkey.2022, Valentini.2025}. However, increasing the amount of ion qubits requires complex processors with high technological demands for the on-chip integration of diverse features \cite{Romaszko2020, Brown.2021, Blain.2021}: i) multi-metal layers and through-substrate vias to enable electrical routing, ii) material and geometry optimization for low dissipation and high heat conduction, iii) integrated optics for laser light routing and ion addressing spanning wavelengths from deep ultra-violet to the infrared depending on the used ion species, iv) integrated detectors for quantum state readout, v) active and passive electronics and optics for on-chip signal multiplexing and manipulation, and vi) deep and stable trapping potentials, optimally using segmented 3D ion traps \cite{Auchter.2022}. Finally, all technologies and materials should be compatible with cryogenic temperature, that allows one to create the extremely high vacuum needed for reliable storage of large numbers of ions. This plethora of needed features leads to a large fabrication complexity and high costs. Combining all features in a monolithic design furthermore creates conflicting requirements, forcing trade-offs that can adversely affect processor performance.

Stacked wafer technology, which interfaces different components, has been developed as a possible mitigation strategy for these technological obstacles \cite{Romaszko2020, Lekitsch.2017}. We extend this idea by proposing a TIQC buildup inspired by the emerging chiplet technology, that uses a system of multiple interfaced chips combined in one package to extend the fabrication possibilities and maximize processor modularity using a divide-and-conquer approach. This chiplet approach shifts complexity from the chiplet fabrication to their packaging. In particular, we propose to realize the ion trap and the integrated optics on individual chiplets, in contrast to state-of-the-art monolithic realizations \cite{Niffenegger.2020, Mehta.2020, Kwon.2024}. A chiplet approach to optics integration simplifies the individual module fabrication and avoids a common compromise in the choice of chip substrates: the manufacturing of complex TIQC processors \cite{Cho.2015,Holz.2020,Auchter.2022, Sterk.2024} usually employs silicon as the basic substrate material due to the low adaption efforts when using standardized complementary metal-oxide-semiconductor (CMOS) processes. However, silicon is not well suited for the application in ion traps, and a lot of fabrication complexity is required to avoid RF and photon absorption within the substrate \cite{Sterk.2024, Dietl.2025}. The fabrication of an ion trap electrode structure on a dielectric substrate material with low RF loss, such as fused silica, crystalline quartz, or sapphire \cite{Bruzewicz.2015,Hite.2012,Daniilidis.2014}, simplifies fabrication efforts once the process technology is developed. Using heterogeneous integration technology \cite{Ramm.2012}, bare ion trap chips with optimal substrate material may then be interfaced with separately fabricated silicon chips carrying standard optical waveguide platforms \cite{Mehta.2016,Niffenegger.2020,Kwon.2024} and CMOS technologies, e.g., photon detection \cite{Todaro.2021,Setzer.2021}, active integrated photonics components \cite{Hogle.2023}, cryogenic amplifier \cite{Sieberer.2021} or digital-to-analog converters (DAC) \cite{Stuart.2019}.

We begin this article with an introduction to the chiplet approach and its benefits for the realization of a scalable TIQC processor. We then demonstrate the realization of a building block for individual ion addressing using chiplet technology, before we conclude by pointing out future prospects for the use of the chiplet approach for TIQC processors.

\section{Chiplet approach for TIQC processors}
\label{sec: chiplet approach}

\begin{figure}[t]
    \centering
    \includegraphics[width=\columnwidth]{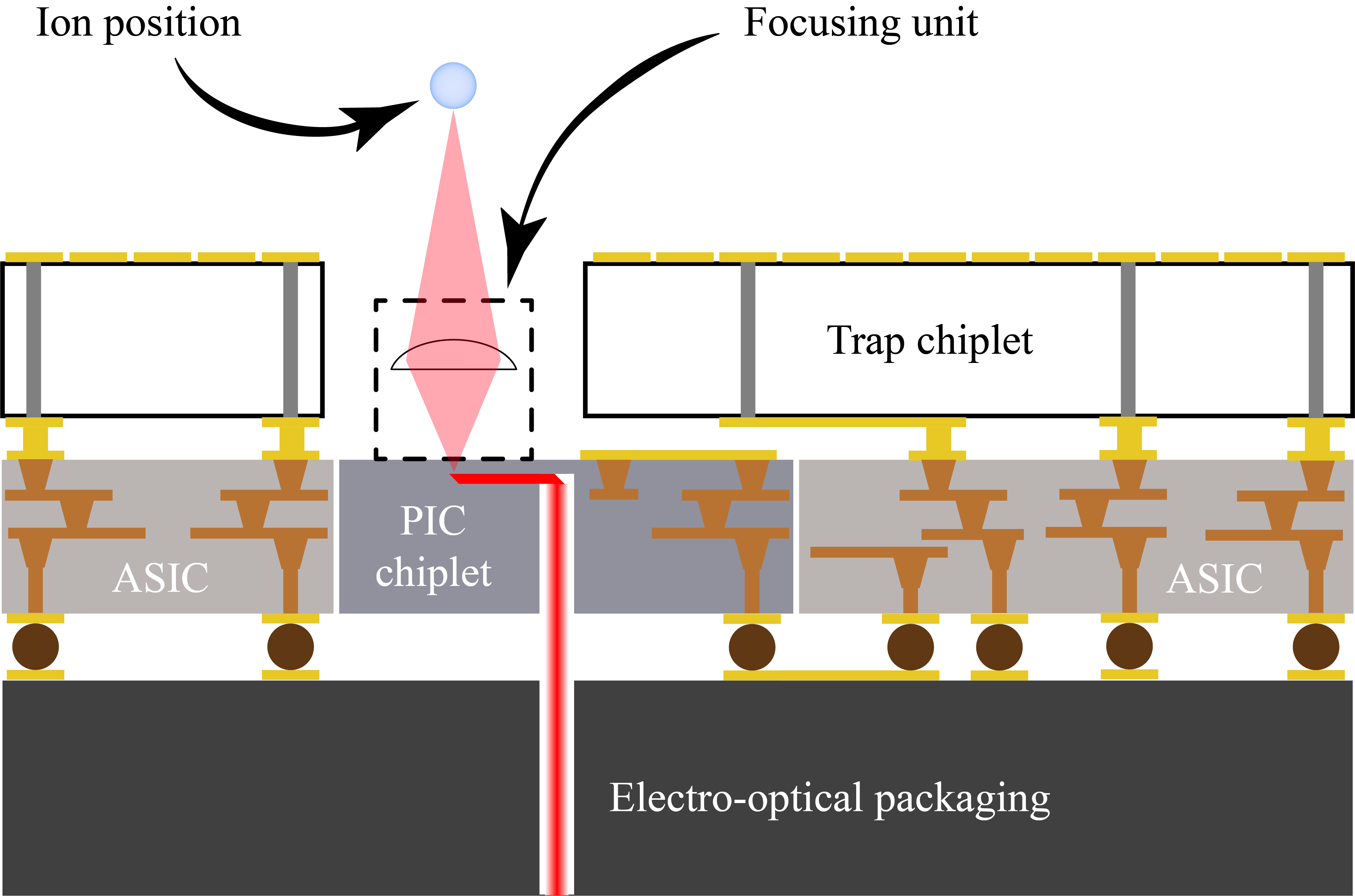}
    \caption{Illustration of a conceptual chiplet stack forming a SiP as platform for scalable TIQC processors. An electro-optical packaging at the bottom serves as an active interposer for the mounted ASICs and PIC chiplets in the middle layer. ASICs and PIC chiplets are laterally interfaced via the interposer and serve in turn as mounting platform for the trap chiplet and further optical elements, e.g., a focusing unit manufactured in a slotted area of the trap chiplet.}
    \label{fig: chiplet_vision}
\end{figure}

In the traditional semiconductor industry, the functionality of a central processing unit (CPU) relies heavily on the density of transistors on a single chip. However, this relationship comes at the cost of increased fabrication complexity and immense processing expenses. Gordon Moore examined the relationship between relative fabrication costs and the number of components per integrated circuit (IC), which is now known as Moore's Law. \cite{Moore.2006,Moore.2006b}. Rather than increasing system complexity through constant feature size reduction, the "More than Moore" approach incorporates additional functionalities, such as RF communication, power control, passive components, sensors, and actuators, on separately fabricated chiplets. The integration of multiple chiplets on a common interposer, along with their interconnection, forms a system-in-package (SiP) \cite{Wang.2023} with a functionality similar to or higher than that possible with monolithic integration alone. 

The "More than Moore" movement \cite{Graef.2021} developed a concept that enables more modular architectures and offers significant technological improvements over monolithic architectures. This concept can be applied to realize a large-scale TIQC processor, where all functionalities are integrated into sevaral chiplets, offering substantial advantages: i) The fusion of multiple manufacturing technologies and materials in a SiP offers the ability to pick the technology that is best suited for different functionalities in the TIQC processor; ii) Parts of the SiP can be revised or modified without having to adapt other system components iii) The modular construction may lead to more reliable sub-components and significantly faster and more cost-efficient development iterations compared to monolithic integrated TIQC solutions, where revisions typically require complex adjustments in the overall manufacturing process. 

Here, we describe a modular SiP architecture for a TIQC processor that consists of multiple chiplets which are placed and interfaced laterally and vertically, as illustrated in Fig.\,\ref{fig: chiplet_vision}. The top layer is reserved for the ion trap (henceforth named trap chiplet), which comprises a substrate material with low RF loss, electrodes on the chiplet top surface for ion trapping, through-substrate vias (TSVs) to connect the electrodes with the chiplet back side, and openings in the substrate for optical access. These openings can be used to host microfabricated optical elements that act as focusing units. The second layer comprises multiple supply chiplets for photonic and electric signal routing and control. Chiplets with photonic integrated circuits (PICs) are used to route and switch optical signals \cite{Hogle.2023}, supplying the laser light for photoionization, laser cooling and qubit control to the ion trapping locations. Chiplets with application-specific integrated circuits (ASICs), for instance integrated DACs \cite{Stuart.2019} or switch-matrices \cite{Malinowski.2023}, are used to route and control the electrical signals for ion trapping and transport operations. The third and bottom layer serves as an active interposer \cite{Coudrain.2019} for the chiplet stack. Besides electrical interfaces commonly used to connect the trap electrodes to electrical supply lines \cite{Stick.2006}, the package may also integrate an optical interface to optical fibers connected to the supply laser, thus forming an electro-optical package. The packaging layer furthermore thermally anchors the chip stack via an attached heat sink.  

Bonding or soldering processes are used to mechanically and  electrically interface the individual chiplets. Possible options encompass metal bonding \cite{Clauberg.2016}, hybrid-bonding \cite{Hsiung.2024}, and solid-liquid interdiffusion (SLID) bonding \cite{SUN.2020}, as well as low-temperature soldering using tin, bismuth, or indium alloys \cite{Illes.2025}. The applicable thermal budget, possible surface preparation, and interconnect density determine which bonding or soldering technology is suitable for SiP chiplet interconnections. The interfacing necessitates its own process developments and plays a key role in the packaging complexity. 

We stress that the chiplet approach, due to its inherently modular nature, goes beyond mere chip stacking. For example, it is challenging to combine PIC and ASIC components monolithically in a single chiplet due to lithography restrictions, contamination issues, or incompatible manufacturing processes. However, such technological limitations can be completely avoided by using separate PIC and ASIC chiplets, as illustrated in Fig.\,\ref{fig: chiplet_vision}, each fabricated using distinct fabrication technologies tailored to the respective chiplet functionality. Likewise, the first SiP layer may comprise several trap chiplets. Furthermore, additional components can be integrated in between two chiplets or in chiplet openings, such as the additional optical elements described in section \ref{sec: realization}.

\subsection{Example realization}

\begin{figure*}
    \centering
    \includegraphics[width=\textwidth]{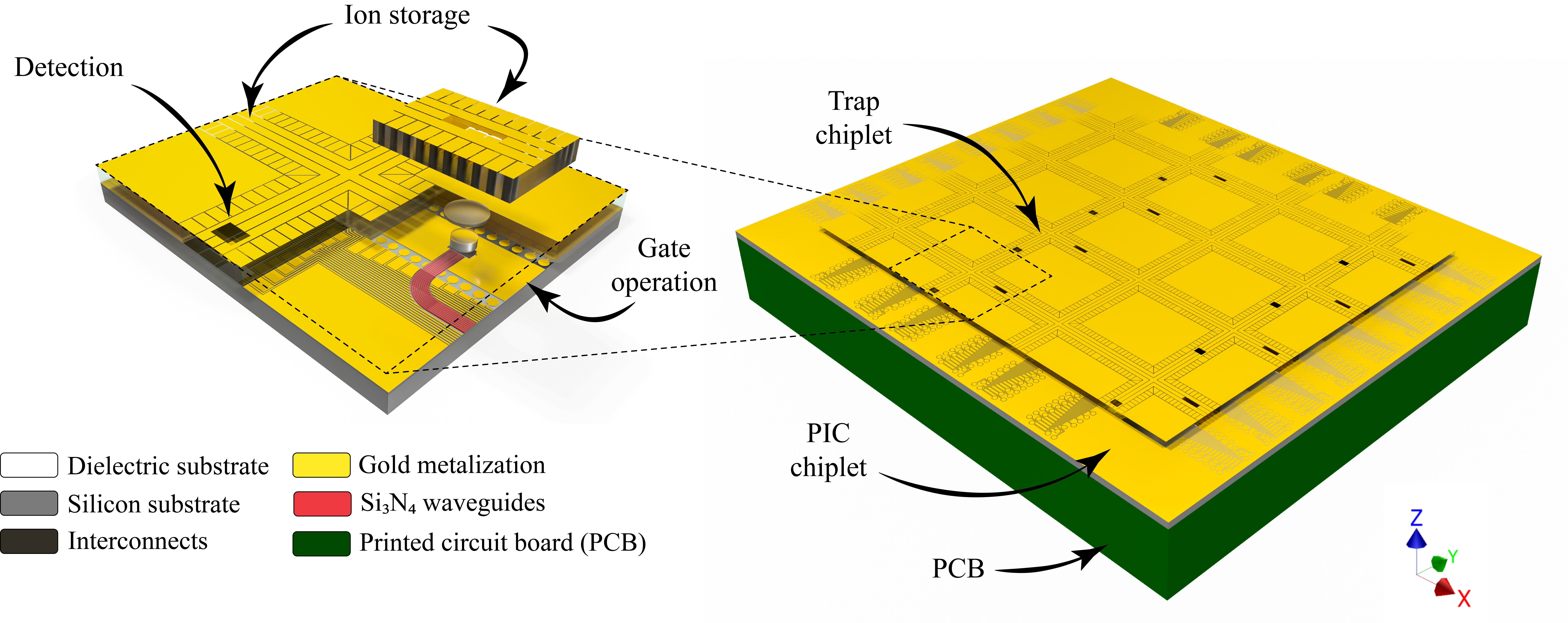}
    \caption{Illustration of a QCCD-type TIQC processor constructed with chiplet technology introduced in Fig.\,\ref{fig: chiplet_vision}. The SiP consists of a trap chiplet bonded onto a PIC chiplet and interfaced on a electro-optical package used for the signal fan-out to peripheral system components. The chiplet stack is mounted via standardized flip-chip bonding technique onto the electro-optical package. Interconnects between dielectric and silicon substrate are created via TSVs and metal-metal bonding technology. The magnified view on a single X-junction on the right shows the parts of the SiP necessary  for the required ion transport and qubit operations in a large-scale TIQC processor. Parts of the trap chiplet are cut off to provide open view on the optical and electrical signal fan-out and ion addressing zone.}
    \label{fig: TIQC_processor}
\end{figure*}

Several architectures for a scalable TIQC processor have been proposed: a quantum charge-coupled device \cite{Kielpinski.2002}, a quantum spring array \cite{Valentini.2025}, processors employing magnetic gradient induced coupling \cite{Khromova.2012} or network approaches using photonic interconnects \cite{Monroe.2014}. For all of these architectures, typically a monolithic integration approach has been considered. However, they can likewise be realized with the chiplet approach, as we will illustrate at the example of a QCCD-type processor. The example TIQC SiP we consider is illustrated in Fig.\,\ref{fig: TIQC_processor}. The ion trap design on the SiP top layer is an array of linear trap sections connected by X-junctions, as envisioned in several previous works, e.g. \cite{Lekitsch.2017, Akhtar.2023}. The magnified view on the left shows a trap region around one of these junctions. Out of the four linear trap "arms" leaving this junction, one arm serves as a laser addressing zone for qubit manipulation using integrated waveguide technology with additional optical elements, while a second arm carries an integrated  photon detector for qubit state detection. The other two arms serve as storage zones and all four arms are connected to adjacent parts of the trap array. A register of ion-qubits can be trapped in such an array using a combination of radio-frequency (RF) and direct current (DC) voltages \cite{Chiaverini.2005} applied to the trap electrodes. The qubit register is thereby distributed over multiple sub-registers. Each sub-register comprises a small number of ions stored in an individual trapping well which can be transported between the different zones of the trap array by adjusting the DC voltages on the trap electrodes. The wells can also be split and merged to enable quantum gate operations between qubits originally in different wells \cite{Pino.2021}.  

As illustrated in the conceptual cross section in Fig.\,\ref{fig: chiplet_vision}, the TIQC-processor in Fig.\,\ref{fig: TIQC_processor} consists of separately manufactured chiplets that are electrically and optically interconnected using heterogeneous integration technology \cite{Ramm.2012}. The trap chiplet carries the RF and DC trap electrodes on a dielectric substrate, which offers beneficial properties with respect to RF loss. Metalized through-substrate vias (TSVs) interconnect the front side of the trap chiplet to its back side. The PIC chiplet on the second layer is fabricated on a silicon substrate to benefit from the standardized production of classical integrated circuits (ICs) and photonic integrated circuits (PICs). In our example realization, the PIC chiplet carries optical waveguides for integrated laser routing and a top metal layer to serve as an additional redistribution layer (RDL) of electrical signals. Microfabricated optical elements are integrated into a trap chiplet trench. These optical elements focus laser radiation emitted from the waveguides onto the ion positions. For this reason, this optical system is named the "focusing unit". 
As depicted in Fig.\,\ref{fig: chiplet_vision}, the second layer may consist of multiple chiplets to increase the functionality of the SiP. Metal or metal-oxide hybrid bonding technologies establish an electrical contact between the trap chiplet back side metal and the PIC chiplet front side metal. The overall chiplet stack is bonded using flip-chip technology on a PCB interposer, which forms the third layer. Optical signals are interconnected by a fiber-to-chip edge coupling technique to the PIC chiplet. Next generation optical interfaces shall be provided through the electro-optical packaging as envisioned in Fig.\,\ref{fig: chiplet_vision}. 

\subsection{Technological advantages}
The realization of a TIQC processor using chiplet technology offers several advantages over monolithic integration with respect to signal routing, power dissipation, and heat management of the processor, and integration of passive and active electronic components enhancing the processor performance. The proposed chiplet technology breaks with the established monolithic integration in that the TIQC processor no longer relies on a specific chip manufacturing technology, e.g., CMOS-compatible semiconductor technology or photonic waveguide technology. This provides significant flexibility in selecting the most appropriate manufacturing techniques for specific components of the TIQC processor and can also speed up development iterations, albeit at the cost of increasing packaging complexity. In general, technologies and realizations can be transferred from monolithically integrated ion traps in a SiP, though this is not always straightforward and can require design changes. 

Specifically, the integration of optical elements for the ion addressing differs from known monolithic techniques that use grating couplers or metasurfaces to focus laser beams \cite{Mehta.2023, Mehta.2016, Niffenegger.2020, Hu.2022}. With the commonly used focusing grating couplers, it is challenging to realize single-ion optical addressing in longer ion crystals \cite{Shirao.2022} due to the large footprint per optical channel. This issue can be solved by using microfabricated lenses, which focus multiple optical channels with the same optical element. Due to the modular nature of the chiplet approach, the integration of such elements is much simpler in a SiP than it would be in using monolithic integration.

Many developed components of monolithic ion traps with integrated optics, e.g. the entire trap electrode design or PIC elements such as optical beam splitters, can be integrated into a SiP without required design changes. In the following subsections, we concentrate on components in a TIQC processor that can be enhanced using the proposed chiplet architecture. 

\subsubsection*{Ion trap substrate}
The separation of the ion trap chiplet in the first layer from the PIC and ASIC supply chiplets in the second layer allows one to select the optimal substrate material for the ion trap chiplet. From a trap performance point of view, the key properties of a good substrate are electrical insulation, low RF loss, large bandgap, and high thermal conductivity. The realization of the trap chiplet on a fused-silica substrate, in combination with TGVs, fulfills these requirements: In contrast to silicon, dielectric substrates, such as fused silica, have superior electric properties because of their low RF loss tangent with minimal heating from RF dissipation in the substrate. Also, due to the large  band gap compared to silicon, laser radiation absorbed in the bulk does not generate mobile charge carriers, which can enhance RF dissipation and may even lead to uncontrolled ion displacement. For these reasons, traps with silicon substrate typically employ additional metal layers to shield the substrate from RF and laser fields \cite{Chung.2025}. Such a multi-metal layer stack, if not carefully designed, leads to high trap capacitance \cite{Sterk.2024}. The omission of the shield layer for dielectric substrates allows the capacitance of the ion trap to be significantly reduced, which in turn reduces the RF dissipation of the trap \cite{Dietl.2025}. 

With respect to the heat management of the ion trap, sapphire or diamond offer high thermal conductivities at cryogenic temperatures, where most ion traps are operated. Additionally, materials with poor heat conductivity, such as fused silica, can be used by exploiting the heat conductivity of the metal filled TGVs which may provide a high effective thermal conductivity from the trap chiplet front side to the heat sink on the packaging back side. 

\subsubsection*{Electrical signal routing}
The electrical signal routing to supply areas is called signal fan-out, and becomes challenging as the TIQC processor scales up to accommodate an increasing number of ions. A single top metal layer cannot provide the required connectivity \cite{Dietl.2025, Bautista-Salvador.2019, Moses.2023}. Thus, multiple stacked and interconnected metal layers, each acting as an RDL for electrical signals, are required. In monolithic fabrication, the signal routing is usually solved by a single-sided multi-metal layer stack. The chiplet realization in Fig.\,\ref{fig: TIQC_processor} provides already two RDLs using the metal layers on the back and front side of the trap and PIC chiplets, respectively. The number of RDLs can be further increased using multi-metal layers on either the trap chiplet back side or on the PIC chiplet front side, without significantly adding to the trap capacitance as long as the RF electrodes are only routed on the trap chiplet top side. With the usage of TGVs, low impedance grounding of DC electrodes in the AC domain could be provided by adding small footprint trench capacitors \cite{Guise.2015, Sterk.2024} to the PIC chiplet, directly below the DC electrodes.

\subsubsection*{Optical signal routing}
A key technological obstacle for large-scale TIQC processors is the integration of optical components that enable the manipulation of the qubit state as well as state readout and laser cooling at the desired zones of the TIQC processor. The state-of-the-art for on-chip laser routing is the application of standardized waveguide technologies \cite{Wörhoff.2015} based on distinct material platforms depending on the specific wavelengths for the used atomic species. All readily available waveguide platforms are based on silicon process technology and offer low-loss laser routing and versatile possibilities for the integration of additional passive \cite{Su.2020} and active \cite{Wang.2025} optical components used for signal multiplexing, mode conversion or suppression, optical switching, spot size conversion and optical filtering. The monolithic integration of waveguide technology on dielectric substrates is principally possible but requires extensive process development to establish a waveguide platform on substrate materials other than silicon with optical performances comparable to those of the available technologies. Due to the high costs for the process development of a waveguide platform on a dielectric substrate, monolithic integration of waveguide technology on dielectric substrates will likely not be available in the near future. Heterogeneous integration technology offers the possibility of using advanced waveguide technology in combination with a surface ion trap fabricated on a dielectric substrate, again with the goal of picking the optimal materials for the individual processor components.

By using a dedicated PIC chiplet for the optical integration of passive and active optical components, the fabrication becomes significantly simpler. The PIC chiplet does not have trap electrodes. With the large separation between silicon material and RF electrodes, and several metal layers in between, the PIC chiplet does not require a shielding layer for the silicon substrate \cite{Doret_2012}. Transparent conductive materials, e.g., indium-titanium oxide (ITO), are an alternative to shield dielectrics in the laser interaction zones of the ion trap \cite{Jansson.2025, Eltony.2013, Niffenegger.2020}. For the chiplet approach, these additional conductive layers can likely be omitted, because dielectrics are not in close proximity to the ions and are already sufficiently shielded by the trap chiplet top metallization. This simplification of fabrication requirements may facilitate the realization of more complex photonic layers with hybrid material platforms routing laser radiation of a broad spectrum from ultra-violet to infrared \cite{Mehta.2023}.

The larger distance between the ion location and the laser out-coupling in the chiplet approach adds the possibility to integrate additional components between the trap chiplet and PIC chiplet to add optical functionality. Examples of this are  printed three-dimensional optical elements which enable individual ion addressing of qubit sub-registers, as we will present in section \ref{sec: realization}.

\subsubsection*{Integrated electronics}
Incorporation of advanced functionalities into a scalable TIQC processor can be achieved through the integration of active electronics, e.g. ASICs, on additional chiplets within the SiP. In the context of DC signal management, the utilization of DACs for signal generation \cite{Stuart.2019} or integrated switch matrices for signal multiplexing \cite{Malinowski.2023} exhibits considerable potential. Likewise, active photonic components need to be integrated to control the laser phase and amplitude for individual optical channels \cite{Hogle.2023}. Regarding the readout of qubit states, the utilization of integrated photodetectors, including single-photon avalanche diodes (SPADs) \cite{Reens.2022} or superconducting nanowire single-photon detectors (SNSPDs) \cite{Hampel.2023}, is a viable option. However, new strategies for qubit state detection can be explored. A promising approach involves the use of grating couplers or microfabricated lenses to couple ion fluorescence in optical channels of the PIC chiplet that are terminated by integrated detectors. This approach may be combined with integrated beam splitters, to generate ion-ion entanglement \cite{Knollmann.2025}.

\subsubsection*{Rapid design modifications and added functionality}
The modular nature of the chiplet approach enables faster and more cost-effective prototyping than monolithic integration allows. Revisions to individual system components do not affect SiP chiplets that are not directly interfaced with the revised parts. A typical revision of the SiP represents a change in trap chiplet electrode design, to realize an optimized surface ion trap architecture without the requirement to adapt the PIC chiplet. Similarly, adaptions of the PIC chiplet, such as the addition of optical channels, a change in photonic routing, or a transfer to a different waveguide platform (e.g. from Si$_3$N$_4$ to Al$_2$O$_3$), do not require an adaptation of the trap chiplet, as the interface remains unchanged. Through this modularity, unrevised SiP components can be fully recycled in the optimized TIQC SiP. New technologies, such as SNSPDs made from high-temperature superconductors \cite{Charaev.2023, Charaev.2024}, can be incorporated into separate chiplets that can be more easily integrated into a functional SiP than a complex, dense integrated chip die. Furthermore, integrating specific functional elements, such as microfabricated lenses, is significantly facilitated with chiplet technology. 

\section{Realization of a chiplet-based TIQC processor}
\label{sec: realization}
We demonstrate the technological advantages of the SiP approach at the example of a TIQC processor building block which enables the optical addressing of single ions in a crystal of ten ions.
We choose this building block because, addressing multiple ions is difficult in monolithically integrated realizations relying on focusing grating couplers. In such a system, each ion requires its own output coupler and thus the footprint becomes prohibitively large such that integrating ten or more independent optical channels next to each other becomes challenging \cite{Shirao.2022}. We circumvent this footprint issue by using a microfabricated lens stack, which uses a single focusing element to address all ions, similar to a macroscopic objective \cite{Pogorelov.2021}. 

The conceptual design of the qubit addressing building block is shown in Fig.\,\ref{fig: Addressing BlBl}a,b. The building block consists of the trap chiplet, which hosts a slotted linear surface ion trap, and is bonded to the PIC chiplet. The slot creates an optical access from the back side of the trap chiplet to the ions, positioned above the RF and DC electrode structure. Individual ions in a linear crystal are addressed with focused laser beams by imaging the output of a waveguide array with a focusing unit consisting of microfabricated lenses placed on top of the PIC chiplet. Each waveguide in the array serves as a dedicated optical channel for one specific ion in the linear ion crystal. The distance between adjacent waveguides corresponds to the ion spacing in a given trapping potential, taking into account the magnification of the focusing unit. The waveguide array terminates at a microfabricated mirror, depicted as metalized ramp, that serves as a deflection element to redirect the laser beams perpendicular to the PIC chiplet surface.
 
\subsubsection*{Ion trap chiplet and electrical routing}
\begin{figure}
    \centering
    \includegraphics[width=\columnwidth]{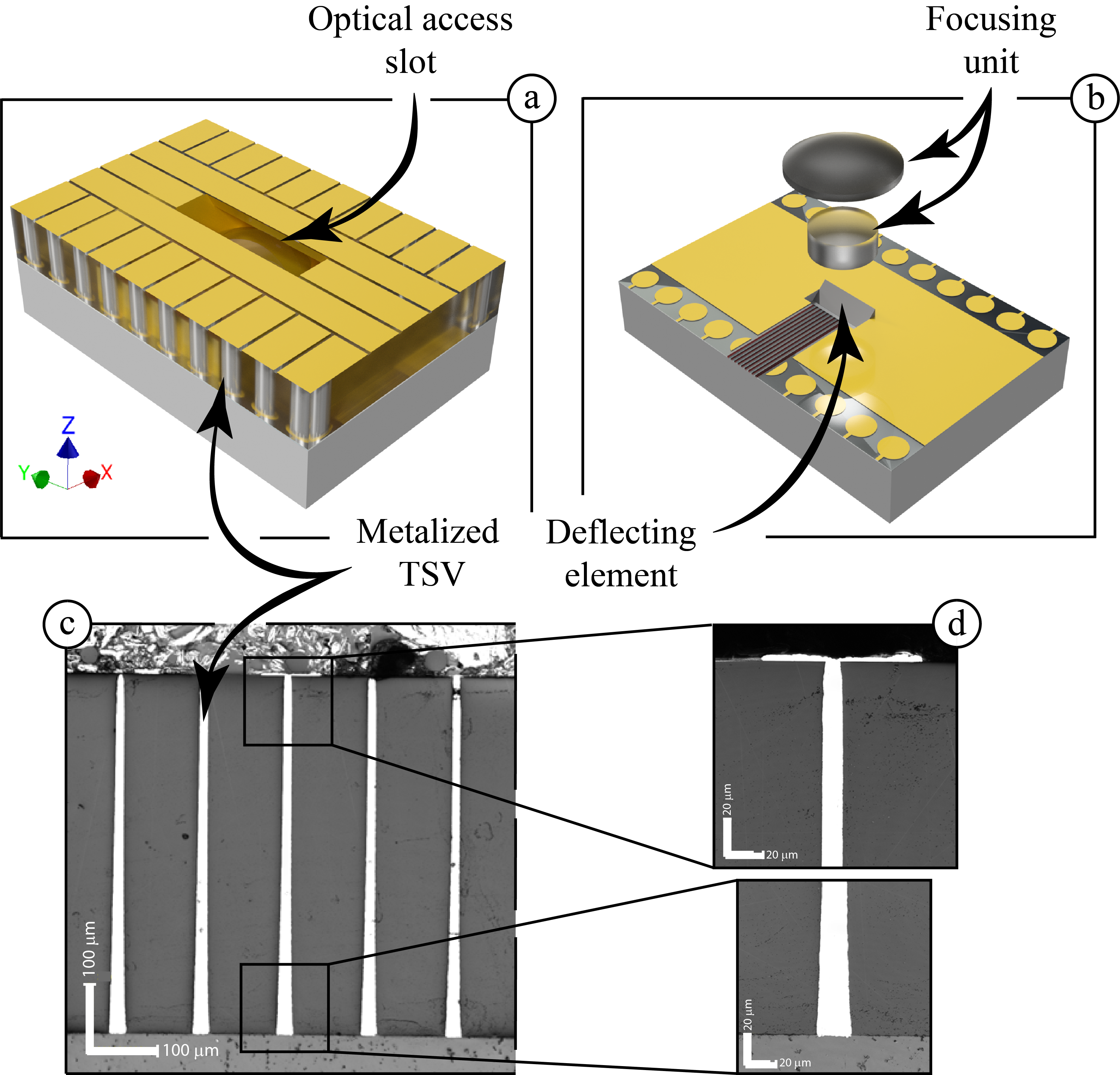}
    \caption{\textcircled{a}: Illustration of a qubit addressing building block, consisting of a slotted trap chiplet with a glass substrate stacked on a PIC chiplet with a silicon substrate. The slot in the trap chiplet  houses an integrated lens stack. \textcircled{b}: Exploded view of the addressing building block without the trap chiplet showing the integrated lens stack that serves as the laser focusing unit. A deflecting element realized on the PIC chiplet redirects the laser beams from waveguides towards the lens stack. \textcircled{c}: Scanning electron microscopy cross section of aluminum filled TSVs electrically interfacing trap chiplet front and back side. \textcircled{d}: Magnified top and bottom part of one TSV, showing no delamination of the aluminum from the substrate after several cool-down cycles to liquid nitrogen temperature.}
    \label{fig: Addressing BlBl}
\end{figure}

For the trap chiplet we employ a borosilicate substrate with gold (Au) electrodes on the front side and an Au routing layer on the back side. We use borofloat glass instead of fused silica due to the better match of the  thermal expansion with the silicon substrate of the PIC chiplet. The electrical connection between the back and front side metal layers on the trap chiplet is realized by TSVs, shown in Fig.\,\ref{fig: Addressing BlBl}c. A magnified image of a single metalized TSV is shown in Fig.\,\ref{fig: Addressing BlBl}d. The demonstrated TSVs consist of high-aspect-ratio cavities produced in the borofloat glass and completely filled with aluminum to ensure electrical conductivity from the back side to the front side metal. The cryo-compatibility of the fabricated TSVs has been tested by cycling the sample structures between ambient and liquid nitrogen temperatures several times. No metal delamination or deformation of the TSV cavities was observed, demonstrating the reliability of the interconnects. We expect that the additional thermal contraction of the TSVs from liquid nitrogen temperature to the operational temperatures of an ion trap around 10 K is negligible \cite{Clark.1968}.   

\subsubsection*{Integrated addressing building block}
\begin{figure*}
    \centering
    \includegraphics[width=0.8\textwidth]{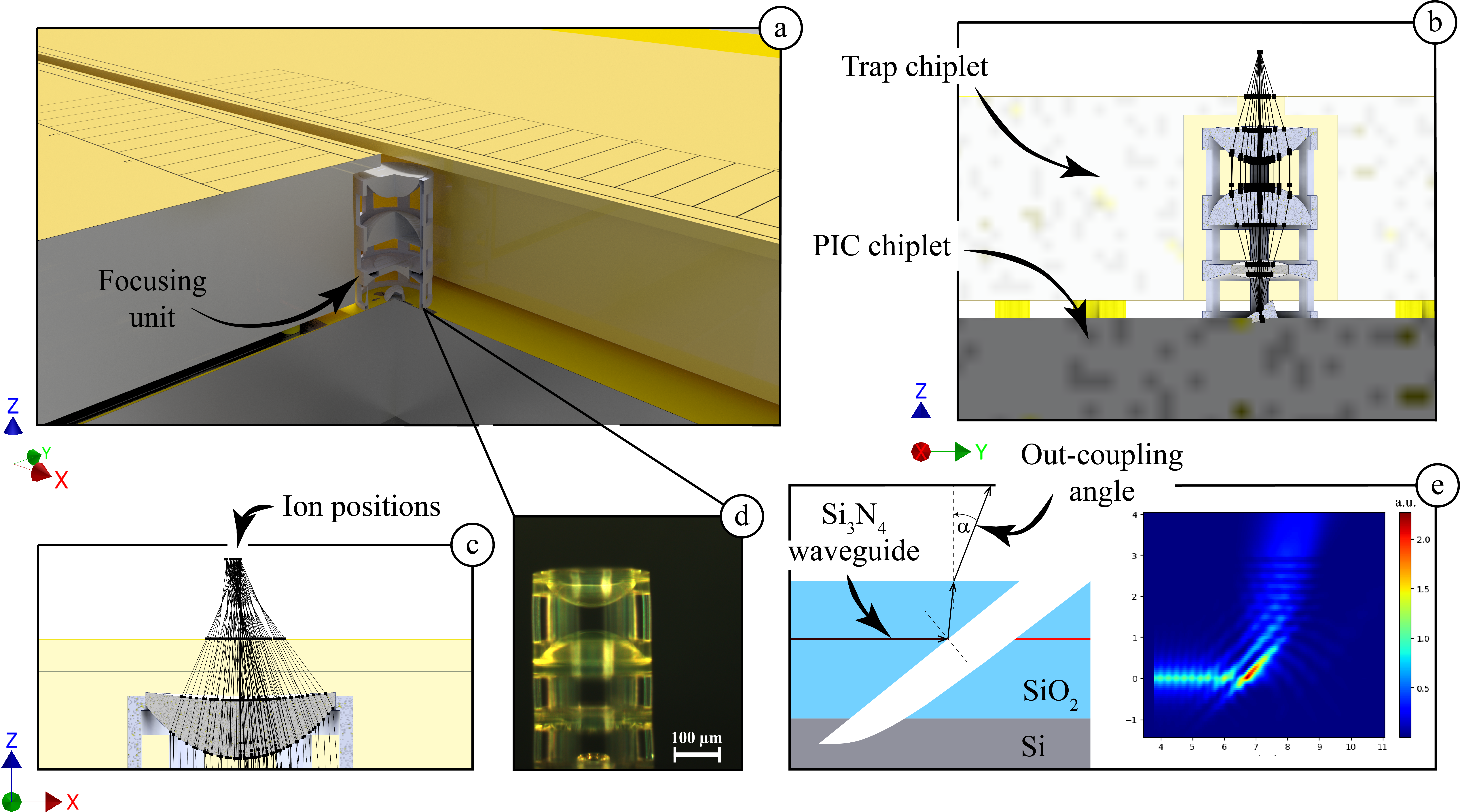}
    \caption{\textcircled{a}: Demonstrator CAD design of the integrated qubit addressing building block. \textcircled{b}, \textcircled{c}: Half-cuts through the CAD design at the focusing unit position in YZ-plane and XZ-plane, respectively. \textcircled{d}: Microscopy image of a fabricated focusing unit on a PIC chiplet. \textcircled{e}(left): Illustration of a TIR mirror structure for a single waveguide and pointing vector of the light propagation. \textcircled{e}(right): Finite-difference time-domain simulation of the light propagation through the TIR mirror structure for a single waveguide.}
    \label{Chips stack design}
\end{figure*}

A more detailed view of the qubit addressing building block is shown in Fig.\,\ref{Chips stack design}a-e. Our demonstrator is designed to allow for individual ion addressing of ten $^{40}$Ca$^{+}$ ions in a linear crystal trapped in a single potential well using 729\,nm light. 
A ten ion crystal has an average ion spacing of about 4\textmu m \cite{James.1998}, which requires highly focused addressing beams for single ion addressing. We use microfabricated optical elements printed directly on the Au-surface of the PIC chiplet using two-photon polymerization technology \cite{OHalloran.2023}. The optical system, depicted in Fig.\,\ref{Chips stack design}b and Fig.\,\ref{Chips stack design}c, is housed in a Au-metalized slot below the laser interaction zone of the trap chiplet and is designed such that each optical channel is imaged onto a corresponding ion position with diffraction limited spot sizes with a numerical aperture of 0.24 and minimized optical crosstalk between adjacent ions. 

The printed focusing unit, shown in Fig.\,\ref{Chips stack design}d, consists of a dielectric material that is likely to induce electric-field noise at the position of the ion crystal, causing motional heating \cite{Teller.2021}. To minimize ion heating, the imaging distance of the optical system is designed to be as large as possible. The metalized and electrically grounded sidewalls of the optical access slot in the trap chiplet additionally shield the ions from stray charges and electric-field noise emanating from the lens stack. 

Optical crosstalk between neighboring optical channels is not only generated by the focusing unit but can also occur on the PIC chiplet due to light leakage from neighboring waveguide structures. Optical simulations determine the crosstalk of adjacent waveguides to approximately \text{-30 dB} for a waveguide distance of 5 \textmu m. The intensity of crosstalk caused by waveguide leakage is similar to the intensity of the simulated crosstalk caused by the focusing unit. For on-chip laser routing at a wavelength of 729 nm, we employ Si$_3$N$_4$ waveguides that are edge-coupled from one PIC chiplet edge. The laser beams in the waveguides are deflected from the PIC chiplet surface plane by total internal reflection (TIR) mirrors, realized directly within the waveguide structures, as shown in Fig.\,\ref{Chips stack design}e(left). The TIR facet within the waveguide array is created by focused ion beam milling a oblique trench into the SiO$_2$/Si$_3$N$_4$/SiO$_2$ layer stack. We employ TIR mirrors instead of the 45\degree ramp depicted in the conceptual illustration, Fig.~\ref{fig: Addressing BlBl}b. 

The laser light guided by the waveguide is reflected upwards at the Si$_3$N$_4$/air TIR-mirror interface and  propagates out of the waveguide cladding at an angle to the surface normal depending on the trench angle. The critical angle of the ion milled trench for TIR in this setup is simulated to be approximately 43\degree.

We fabricate the TIR mirrors at an angle of 52\degree, resulting in an out-coupling angle $\alpha\approx 20\degree$ as this angle gives the highest fabrication accuracy. 
A finite-difference time-domain simulation of the laser propagation at the TIR mirror structure depicted in Fig.\,\ref{Chips stack design}e(right) shows the out-coupled intensity distribution. We use a prism structure, indicated in Fig.~\ref{Chips stack design}b, printed below the first lens of the focusing unit to correct the light propagation and achieve a perpendicular out-coupling out of the PIC chiplet, minimizing optical aberrations in the focusing unit. The other optical components of the focusing unit are designed to generate elliptical focal spots with high focusing ability in the x-direction and a larger spot size in the y-direction. This elliptical beam shape, with aspect ratio of one two three, ensures minimal optical crosstalk between neighboring ions, but relaxes the requirements for chip alignment accuracy. A dark field microscopy image of the fabricated focusing unit printed onto the TIR mirror position is shown in Fig.\,\ref{Chips stack design}d. 

\begin{figure}
    \centering
    \includegraphics[width=\columnwidth]{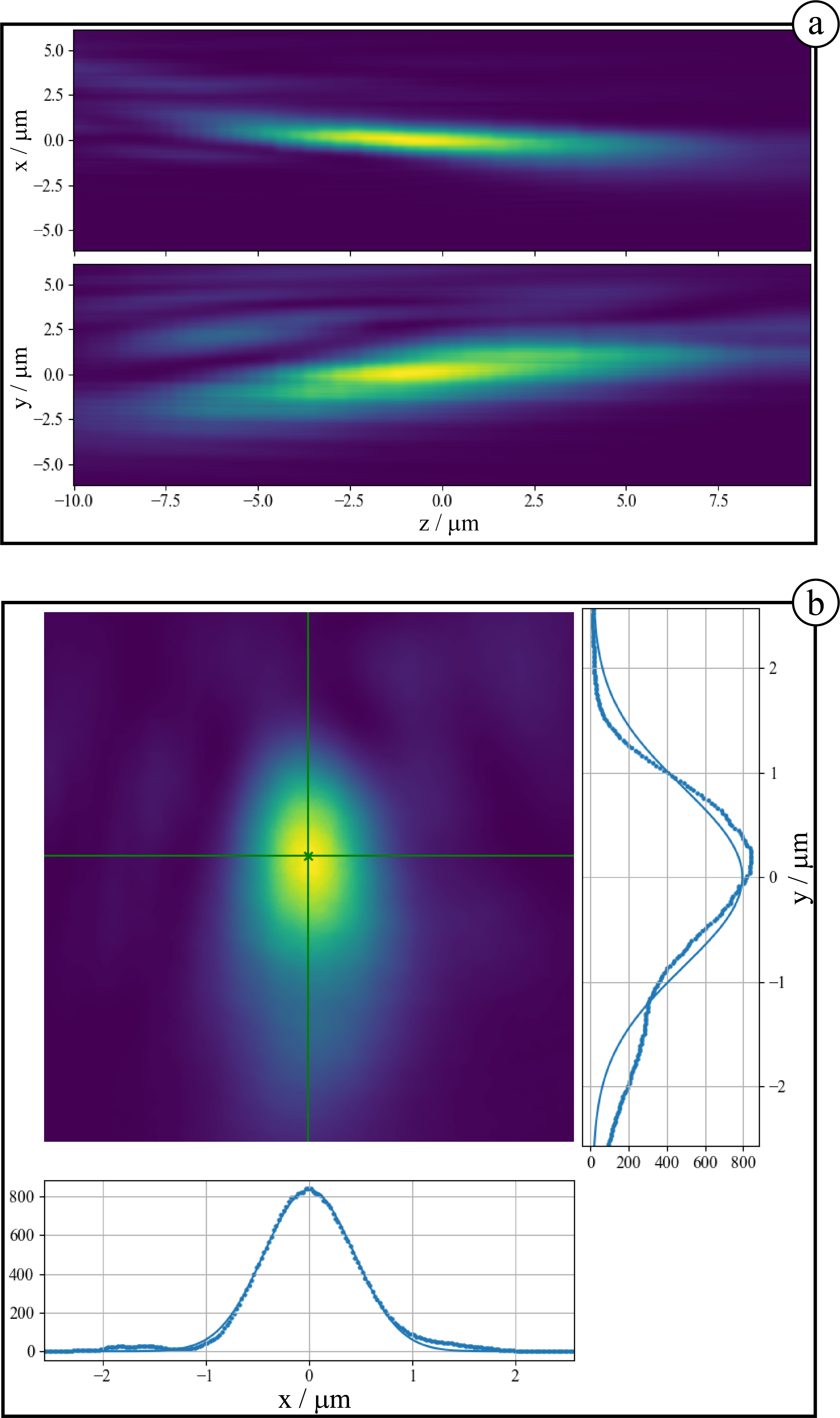}
    \caption{\textcircled{a}: Beam profile of one of the optical channels of the integrated addressing unit showing the light propagation in x- and y-direction around the focal spot. \textcircled{b}: 2D plot of the focal spot at the ion location and 1D slices in x- and y-direction with Gaussian fit functions. The mode field diameter (MFD) in x- and y-direction at the focal spot is determined with MFD$_x$\,=\,(1.73$\pm$0.01) \textmu m and MFD$_y$\,=\,(3.42$\pm$0.02) \textmu m, respectively.}
    \label{fig:Measurements focusing unit}
\end{figure}

We have tested the optical performance of the Si$_3$N$_4$ waveguide design, the TIR mirrors, and the focusing unit with separately fabricated PIC dies equipped with focusing units identical to the one shown in Fig.\,\ref{Chips stack design}d. The optical system is designed to have a magnification of 0.6, which determines the required waveguide spacing when addressing  a linear crystal of ten $^{40}$Ca$^{+}$ ions. At an axial common mode frequency of the ion crystal of 700 MHz, the required waveguide spacing varies between (4.84-6.62) \textmu m.  We determine the magnification of the optical system by balancing optical aberrations originating from the lens stack with crosstalk between adjacent waveguides. We characterize the performance of the building block by measuring spot sizes and focal distances. We couple light into the waveguides using a 6-axis translation stage that holds the PIC chiplet with the printed focusing unit and characterize the spot size of the outcoupled light with a microscope. A typical measurement of a waveguide that illuminates the third outer ion is shown in Fig.\,\ref{fig:Measurements focusing unit}a,b.    

The measurements show a focal spot size in the x-direction of 1.7 \textmu m, and 3.4 \textmu m in the y-direction. The focal spot is measured at a distance of (169\,$\pm$\,6) \textmu m above the focusing unit, which differs from the simulated value by 8 \textmu m. Taking the Rayleigh length of approximately \text{13 \textmu m} into account, the focal spot size and distance correspond well to the simulated data. The simulation yields a mode-field diameter (MFD) at a focal distance of \text{177 \textmu m} in x- and y-direction to be 1.3 \textmu m and 3.5 \textmu m, respectively. We note that the light propagation angle in the y-direction shows a deviation from the intended propagation normal to the PIC surface. This deviation originates from an incorrectly shaped prism structure below the lens stack which was designed for a TIR angle of 7° while the actual out-coupling angle is 20°.  This deviation from the model causes light clipping on the focusing unit in the y-direction. Through further optimization, we are confident that we can match the optical design to the actual out-coupling in an improved version.   

\subsubsection*{Chiplet bonding}
\begin{figure*}
    \centering
    \includegraphics[width=\textwidth]{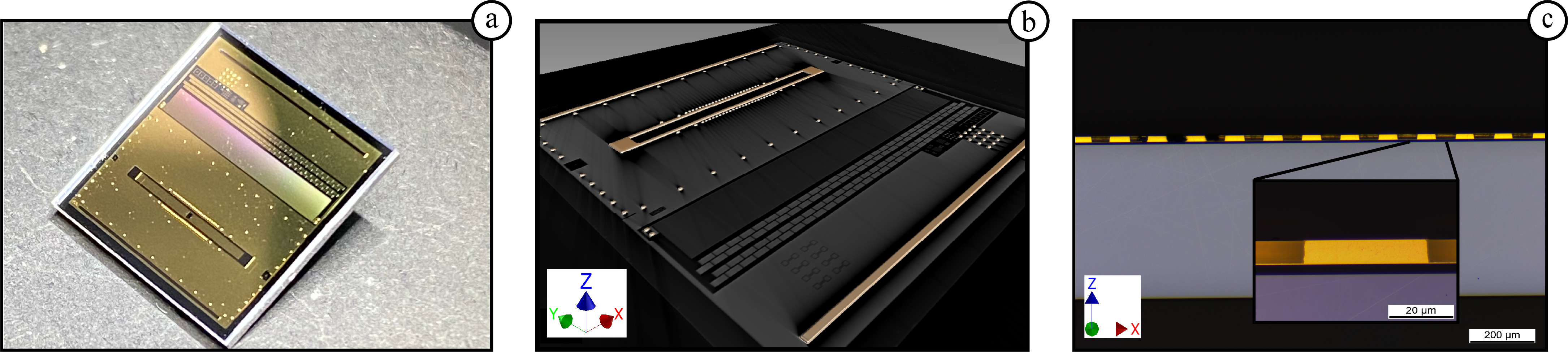}
    \caption{\textcircled{a}: Image of a fabricated chiplet assembly. The trap chiplet metallization is a simplified design with only the bond structures required for this bond demonstration. \textcircled{b}: CT scan of the bond structures on top of the PIC chiplet metalization. \textcircled{c}: Micrograph image of several bonding interfaces with an enlarged micrograph image of a single bond structure.}
    \label{fig:Bonding}
\end{figure*}

Finally, we demonstrate the bonding of the trap chiplet to the PIC chiplet, which is a crucial step in fabricating the proposed TIQC SiP. The thermal expansion of the trap and the PIC chiplet substrate materials differs, which causes stress and makes high-temperature bonding technologies unsuitable for interfacial bonding. We therefore employ a low-temperature thermo-compression bond technique. We additionally compensate for the stress resulting from the thermal expansion mismatch of the substrate materials by employing a thick Au-bonding interface of about 20 \textmu m. Typical metal-metal bonding requires a maximum surface roughness of approximately (2-3) nm$_{\text{rms}}$. With a thick galvanic gold layer, this surface roughness specification is the key factor in achieving high-quality bonding at low temperatures. The maximum heat limit during bonding is determined by the polymer material of the focusing unit. Since overheating can cause damage, it's important to keep the bonding temperature below the point at which the carbon compounds in the polymer begin to evaporate. In collaboration with EV Group GmbH, LioniX International BV and Printoptix GmbH we developed an appropriate thick Au metal-metal bonding process compatible with these restrictions. A bonded chiplet stack with a simplified version of the addressing building block without slots in the trap chiplet and without printed lenses is shown in Fig.\,\ref{fig:Bonding}a-c. 
The bonding process was developed on  wafer scale using a 100-mm silicon wafer and a 200-mm borosilicate wafer. The silicon wafer was manufactured using standard silicon material with a Si$\text{O}_2$ layer of 8 \textmu m thickness and a 500-nm thick structured Au layer. The borosilicate wafer was patterned with 20 \textmu m thick Au structures only on its back side. Figure\,\ref{fig:Bonding}a shows an image of bonded chiplet stack after dicing. In Fig.\,\ref{fig:Bonding}b, an x-ray computer tomography (CT) image of the thick Au bonding structures, with a resolution limit of 7 \textmu m, is shown. Micrograph images from the bonding interface are shown in Fig.\,\ref{fig:Bonding}c. On the tested samples, we cannot observe any open interfaces. The zoom-in image of a single bonding structure in Fig.\,\ref{fig:Bonding}c proves an excellent interface between the Au-surfaces, so we expect a high electrical conductivity of these structures, to be verified in upcoming electrical measurements. Several additional CT images and an acoustic microscopy image of the sealing structure are shown in Appendix \ref{Ap:CT measurements}. These additional investigations revealed no bonding failures in any of the imaged structures. A defect-free bond is especially important for the sealing structure, visible as rectangle around the chiplet slot, which is intended to enclose the focusing unit in the final chip stack to protect it from water and residues produced during the dicing of the wafer stack. The wedge error between the wafers was not measurable within the measurement resolution of 0.002\degree.

\section{Conclusion and Outlook}

The chiplet architecture presented in this work extends the established paradigm of TIQC processor manufacturing, based on monolithic integration of the optical and electrical components, to a SiP approach. We have presented how in a SiP, different functional elements of the processor are realized on multiple chiplets that are optically and electrically interconnected. We have further discussed the technological advantages of the SiP approach. In particular, we discussed how it combines well-established processes for optical and electrical elements with fewer adaptations. The SiP approach also allows for greater flexibility in material selection and opens the door to novel functional elements that are difficult to realize with monolithic integration.
Compared to monolithic integration, the SiP approach reduces the fabrication complexity of the individual chiplets, at the price of added complexity in the packaging of the SiP. The SiP approach makes design iterations and development of new functionalities faster and more cost-efficient, once the chiplet interfacing is established and remains unchanged. 

We have described a proof-of-concept realization of a chiplet-based TIQC processor, which incorporates an integrated laser-addressing building block suited for individual ion addressing of ten ions in a linear crystal at a wavelength of 729 nm. The fabricated chiplet stack for this processor consists of the trap and PIC chiplet which are interfaced via Au-metal bonding. The trap chiplet is realized on a borosilicate glass substrate and carries aluminum filled TGVs for electrical interconnecting front metalization with the trap chiplet back side and PIC chiplet front side metalization used for signal fan-out. The PIC chiplet is equipped with optical waveguides and TIR mirrors for laser routing and out-coupling realized on a silicon substrate material. For individual ion addressing, a 3D-printed lens stack for imaging a waveguide array onto the ion positions is fabricated in a trench of the trap chiplet. The optical characterization of the addressing building block showed diffraction-limited spot sizes of about 1.7 \textmu m in the trap's axial direction at a focal length of about 169 \textmu m, consistent with simulations. We further demonstrated successful bonding of the trap chiplet to the PIC chiplet using a low-temperature thermo-compression bond with an Au bond interface of about 20 \textmu m thickness, which mitigates stress from the thermal expansion mismatch of the two chiplets. The demonstrated wafer-to-wafer bonding process was performed with relatively high forces to compensate for an Au surface roughness of approximately (8-14) nm. We assume that a die-to-die version of the bonding process requires a further reduction in the roughness of the gold surface. A die-to-die bond would be beneficial due to the simplification of the packaging workflow where the PIC chip facets could be polished prior to the bonding using established procedures.

In the future, the integrated addressing building block can be supplemented by additional integrated optics to also provide laser beams for ionization, laser cooling, and state readout. For these beams, individual ion addressing is typically not required such that the printed lens stack could be significantly simplified and its footprint reduced. The focusing unit could be further improved using an achromatic 3D printed lens stack, where polymers with different refractive index are being combined \cite{Schmid.2021}. This could further reduce aberrations and allow one to use multiple addressing beams at different wavelengths to be focused by the same lens stack. 3D-printed polymer lenses will likely show problems for near-UV wavelengths, where currently employed polymers have high optical absorption and may even undergo bleaching. A hybrid approach, combining the established focusing grating couplers \cite{Mehta.2016, Niffenegger.2020} for global beams with the integrated addressing unit shown here, may thus be beneficial. The beams emitted from grating couplers could also propagate through the transparent substrate material of the trap chiplet, removing the need for additional optical access slots.

Due to its inherent modular nature, chiplet integration is also ideal for incorporating additional technologies into large-scale TIQC processors. Integrated electronics such as digital-to-analog converters \cite{Stuart.2019} or electrical switching electronics \cite{Malinowski.2023} could be realized on the PIC chiplet where the silicon substrate offers compatibility with established CMOS technologies. Similarly, passive components such as trench capacitors \cite{Sterk.2024} for RF grounding of the trap's DC control electrodes could be added. Other technologies may comprise active PIC elements to control laser phase, frequency, and amplitude \cite{Hogle.2023}, but also CMOS-compatible microheaters serving as a neutral atom source for ion loading \cite{Kumar.3132025}. The latter would require a loading slot in the trap chiplet, similar to the optical access slot shown in this work. Integrated photon detectors \cite{Todaro.2021, Reens.2022} may likewise be added to the PIC chiplet. Here, one could even combine a 3D-printed focusing unit similar to the one shown in this work with a detector array to facilitate simultaneous state-readout of multiple ions in a crystal. Alternatively, the fluorescence light of individual ions in a chain could be coupled in an array of waveguides, each terminating at a detector, to further reduce readout crosstalk or allowing for photon-detection-mediated entangling operations \cite{Knollmann.2024, Knollmann.2025}. We stress that all these technologies do not need to be realized on a single chiplet. Multiple chiplets with different functional elements may be combined, e.g., laterally placed next to each other on a "supply layer" below the trap chiplet. 

Finally, the SiP approach also offers a viable technological solution for the heterogeneous integration of optical components into segmented 3D ion traps produced by selective laser-induced etching (SLE) \cite{Simeth.1282023232023}. Since these traps typically employ glass substrates, one could possibly directly transfer the developed Au-metal bonding technology developed in this work to interface a segmented 3D ion trap to a PIC chiplet.  

\begin{acknowledgments}
\label{acknow}
The authors thank Dr. Simon Thiele of Printoptix GmbH for helpful discussions on two-photon polymerization technology and Dr. Renilkumar Mudachathi and Raimond Frentrop for their offered knowledge on the LioniX International BV waveguide technology. We would also like to thank Thomas Stöttinger and Marcus Dornetshumer of EV Group GmbH for sharing their knowledge in bonding technologies and for their willingness to work outside of standard specification guidelines. We gratefully acknowledge support by the European Union’s Horizon Europe research and innovation program under Grant Agreement Number 101114305 (“MILLENION-SGA1” EU Project), the European Union’s Horizon Europe research and innovation program under Grant Agreement Number 101046968 (BRISQ), the Austrian Science Fund (FWF Grant-DOI 10.55776/F71, 10.55776/COE1), the Austrian Research Promotion Agency (FFG) under the project “ScaleQudits” (914032), "SIQCI" (891364), "ITAQC" (896213), and "GLASTRAP" (915438), as well as the Intelligence Advanced Research Projects Activity (IARPA) and the Army Research Office, under the Entangled Logical Qubits program through Cooperative Agreement Number W911NF-23-2-0216. 

\section*{Author Contributions}
B.B., P.H., T.M., and P.S. designed the experiment. B.B., M.R., and N.J. carried out the measurements and analyzed the data. B.B., P.H., H.Q., and D.K. fabricated the devices. B.B., P.H., and P.S. wrote the manuscript. All authors reviewed the manuscript. J.U., T.M., and P.S. supervised the project.

\end{acknowledgments}
\appendix

\section{Additional bonding investigations}
\label{Ap:CT measurements} 
\begin{figure}[ht]
    \centering
    \includegraphics[width=\columnwidth]{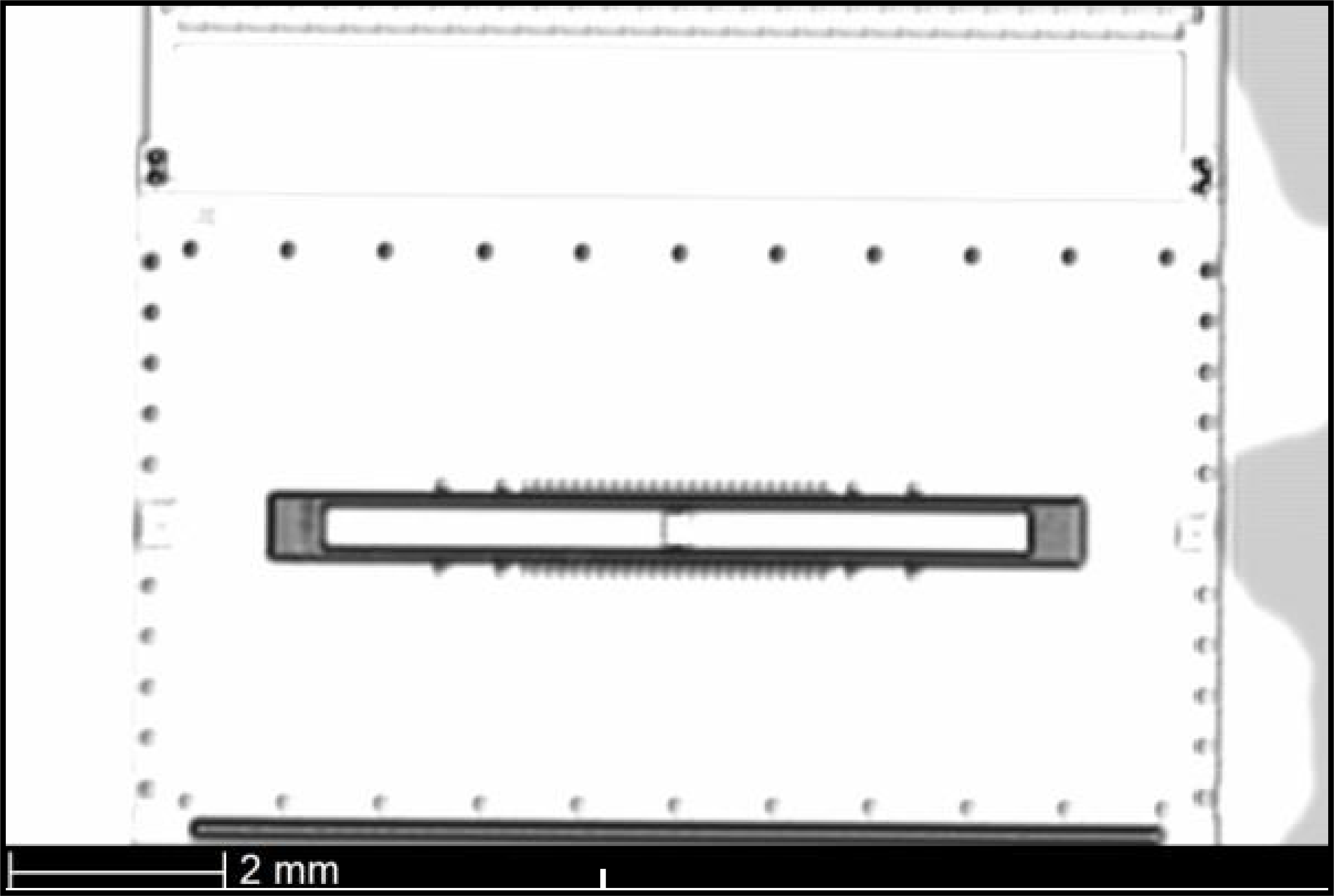}
    \caption{Acoustic microscopy image of the sealing structure and the surrounding bonding pads. The dark areas indicate solid contact at the bonding surface. White areas represent open spaces. The absence of dark areas within the sealing structure indicates proper sealing function and the absence of water.}
    \label{fig:Acoustic_microscopy}
\end{figure}

\begin{figure}[ht]
    \centering
    \includegraphics[width=\columnwidth]{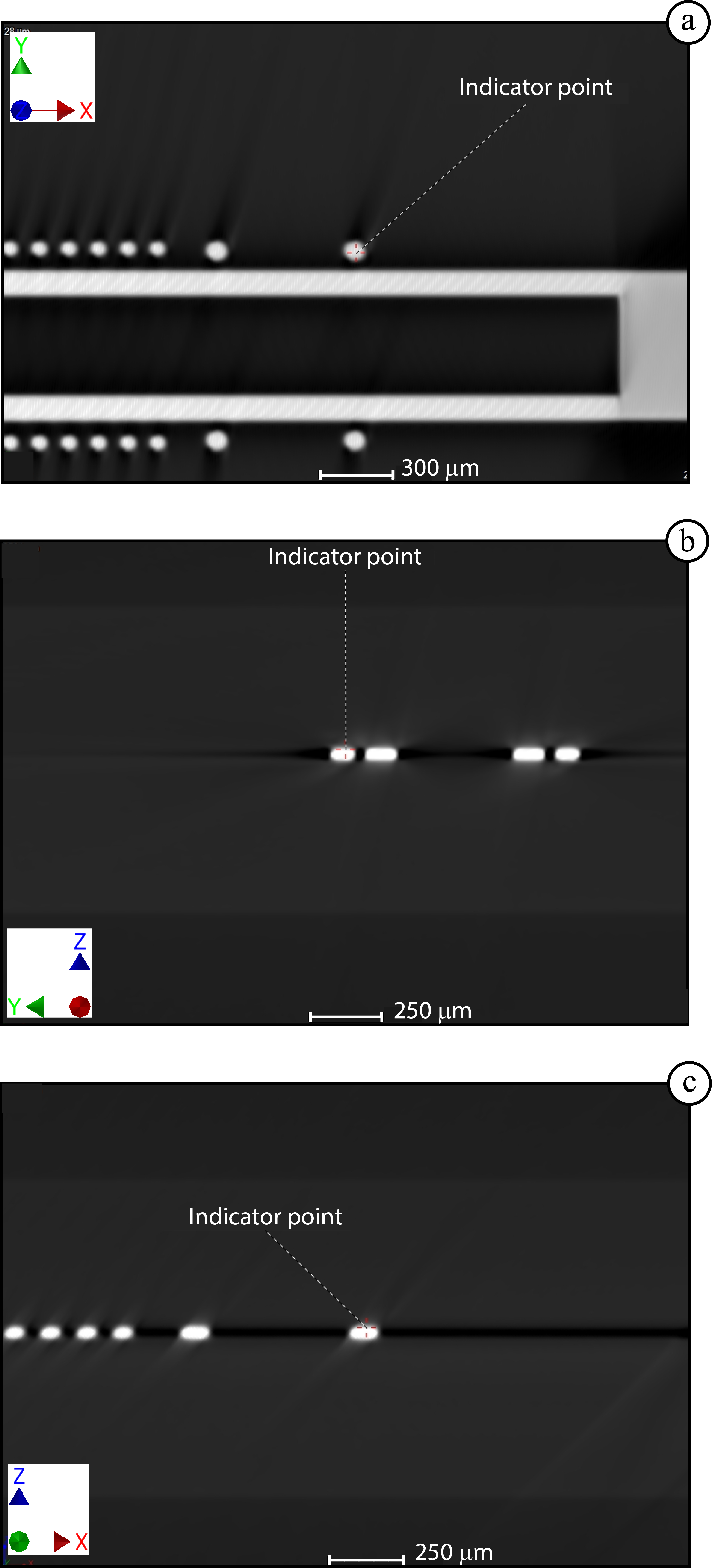}
    \caption{\textcircled{a}: CT scan of the bonding interface with the rectangular sealing structure around the chipet slot and multiple bonding pads (round) for electrical supply to the DC electrodes via the chiplet TGVs. White areas mark regions of high density (i.e. metal). \textcircled{b}: Cross section of the sealing and electrical bonding pads at the indicator point in the YZ-plane. \textcircled{c}: Cross section of the sealing and electrical bonding pads at the indicator point in the XZ-plane.}
    \label{fig:CT_measures}
\end{figure}

We verify the quality of the wafer-to-wafer Au-metal bond by performing CT measurements of the bond structures. We examine the bonding interfaces for open areas.
in the Au-metal to ensure proper bonding quality, which compensates for the mechanical stress produced by the different thermal expansions of the trap and PIC chiplet substrate materials. This also ensures good electrical conductivity for routing the ion trap signals and seals the printed micro lens stack of the focusing unit during dicing. CT measurements in Fig.\,\ref{fig:CT_measures}a and the acoustic microscopy investigations shown in Fig.\,\ref{fig:Acoustic_microscopy} of an example sealing structure reveal an enclosed bonding surface throughout the observed interface. Additionally, no open metal areas were observed on the bonding pads surrounding the sealing structure. The cross sections in YZ-plane and XZ-plane in Fig.\,\ref{fig:CT_measures}b,c confirm the successful bonding process.

In addition to the bonding quality, the wedge error between the trap and PIC chiplets was investigated. This is crucial for the function of the ion addressing building block, in which highly focused laser beams illuminate a linear ion crystal. If the crystal were not parallel to the PIC chiplet surface, it could not be guaranteed that the ions would see the smallest spot size. This would increase the optical crosstalk between several channels. We digitally measured the wedge error between both chiplets with cross sections of the chiplet stack in the XZ and YZ planes using CT scans. The error was determined to be 0.000\degree$\pm$0.002\degree.

\FloatBarrier
\bibliography{bibliography}

@misc{Hughes.2025,
      title={Trapped-ion two-qubit gates with $>$99.99{\%} fidelity without ground-state cooling}, 
      author={A. C. Hughes and R. Srinivas and C. M. Löschnauer and H. M. Knaack and R. Matt and C. J. Ballance and M. Malinowski and T. P. Harty and R. T. Sutherland},
      year={2025},
      eprint={2510.17286},
      archivePrefix={arXiv},
}

@article{Pogorelov.2021,
  title = {Compact Ion-Trap Quantum Computing Demonstrator},
  author = {Pogorelov, I. and Feldker, T. and Marciniak, Ch. D. and Postler, L. and Jacob, G. and Krieglsteiner, O. and Podlesnic, V. and Meth, M. and Negnevitsky, V. and Stadler, M. and H\"ofer, B. and W\"achter, C. and Lakhmanskiy, K. and Blatt, R. and Schindler, P. and Monz, T.},
  journal = {PRX Quantum},
  volume = {2},
  issue = {2},
  pages = {020343},
  numpages = {23},
  year = {2021},
  month = {Jun},
  publisher = {American Physical Society},
  doi = {10.1103/PRXQuantum.2.020343},
  url = {https://link.aps.org/doi/10.1103/PRXQuantum.2.020343}
}

@article{Eltony.2013,
    author = {Eltony, Amira M. and Wang, Shannon X. and Akselrod, Gleb M. and Herskind, Peter F. and Chuang, Isaac L.},
    title = {Transparent ion trap with integrated photodetector},
    journal = {Appl. Phys. Lett.},
    volume = {102},
    number = {5},
    pages = {054106},
    year = {2013},
    month = {02},
    issn = {0003-6951},
    doi = {10.1063/1.4790843},
    url = {https://doi.org/10.1063/1.4790843},
}

@article{Jansson.2025,
author = {Erik Jansson and Volker Scheuer and Elena Jordan and Konstantina Kostourou and Tanja E. Mehlst\"{a}ubler},
journal = {Appl. Opt.},
number = {7},
pages = {1715--1722},
publisher = {Optica Publishing Group},
title = {Indium tin oxide combined with anti-reflective coatings with high transmittance for wavelengths <400 nm},
volume = {64},
month = {Mar},
year = {2025},
url = {https://opg.optica.org/ao/abstract.cfm?URI=ao-64-7-1715},
doi = {10.1364/AO.547471},
}

@article{Chung.2025,
doi = {10.1088/2058-9565/add04c},
url = {https://doi.org/10.1088/2058-9565/add04c},
year = {2025},
month = {may},
publisher = {IOP Publishing},
volume = {10},
number = {3},
pages = {035014},
author = {Chung, Daun and Choi, Kwangyeul and Lee, Woojun and Kim, Chiyoon and Shon, Hosung and Park, Jeonghyun and Cho, Beomgeun and Lee, Kyungmin and Kim, Suhan and Yoo, Seungwoo and Jung, Uihwan and Jung, Changhyun and Kang, Jiyong and Kim, Kyunghye and Berkis, Roberts and Northup, Tracy and ‘Dan’ Cho, Dong-Il and Kim, Taehyun},
title = {A silicon-based ion trap chip protected from semiconductor charging},
journal = {JQST}
}

@article{Shirao.2022,
doi = {10.35848/1347-4065/ac5b27 },
url = {https://doi.org/10.35848/1347-4065/ac5b27       } ,
year = {2022},
month = {jun},
publisher = {IOP Publishing},
volume = {61},
number = {SK},
pages = {SK1002},
author = {Shirao, Mizuki and Klawson, Daniel and Mouradian, Sara and Wu, Ming C.},
title = {High efficiency focusing double-etched SiN grating coupler for trapped ion qubit manipulation},
journal = { Jpn. J. Appl. Phys.}
}

@article{Doret_2012,
doi = {10.1088/1367-2630/14/7/073012 },
url = {https://doi.org/10.1088/1367-2630/1   4/7/073012     }   , 
year = {2012},
month = {jul},
publisher = {IOP Publishing},
volume = {14},
number = {7},
pages = {073012},
author = {Charles Doret, S and Amini, Jason M and Wright, Kenneth and Volin, Curtis and Killian, Tyler and Ozakin, Arkadas and Denison, Douglas and Hayden, Harley and Pai, C-S and Slusher, Richart E and Harter, Alexa W},
title = {Controlling trapping potentials and stray electric fields in a microfabricated ion trap through design and compensation},
journal = {New J. Phys.}
}

@article{Wang.2025,
 author = {Wang, Hui and Ralph, Timothy C. and Renema, Jelmer J. and Lu, Chao-Yang and Pan, Jian-Wei},
 year = {2025},
 title = {Scalable photonic quantum technologies},
 journal = {Nature materials},
 doi = {10.1038/s41563-025-02306-7 }
}

@article{Su.2020, 
author = {Su, Yikai and Zhang, Yong and Qiu, Ciyuan and Guo, Xuhan and Sun, Lu},
title = {Silicon Photonic Platform for Passive Waveguide Devices: Materials, Fabrication, and Applications},
journal = {Adv. Mater. Technol.},
volume = {5},
number = {8},
pages = {1901153},
doi = {https://doi.org/10.1002/admt.201901153           } ,
url = {https://advanced.onlinelibrary.wiley.com/doi/abs/10.1002/admt.201901153},
year = {2020}
}

@article{Wörhoff.2015,
url = {https://doi.org/10.1515/aot-2015-0016}                ,
title = {TriPleX: a versatile dielectric photonic platform},
author = {Kerstin Wörhoff and René G. Heideman and Arne Leinse and Marcel Hoekman},
pages = {189--207},
volume = {4},
number = {2},
journal = {Adv. Opt. Technol.},
doi = {doi:10.1515/aot-2015-0016}                ,
year = {2015},
lastchecked = {2025-10-28}
}

@INPROCEEDINGS{Illes.2025,
  author={Illés, Balázs and Géczy, Attila and Tafferner, Zoltán and Skwarek, Agata and Krammer, Olivér},
  booktitle={2025 ISSE}, 
  title={Low-Temperature Soldering (LTS) in the Electronics Industry: a Brief Review}, 
  year={2025},
  volume={},
  number={},
  pages={1-6},
  doi={10.1109/ISSE65583.2025.11120985}}

@INPROCEEDINGS{Clauberg.2016,
  author={Clauberg, Horst and Rezvani, Alireza and Venkatesan, Vinod and Frick, Guy and Chylak, Bob and Strothmann, Tom},
  booktitle={2016 IEEE 66th ECTC}, 
  title={Chip-to-Chip and Chip-to-Wafer Thermocompression Flip Chip Bonding}, 
  year={2016},
  volume={},
  number={},
  pages={600-605},
  doi={10.1109/ECTC.2016.329}}

@ARTICLE{Hsiung.2024,
  author={Hsiung, Chien-Kang and Chen, Kuan-Neng},
  journal={IEEE Nanotechnol. Mag.}, 
  title={A Review on Hybrid Bonding Interconnection and Its Characterization}, 
  year={2024},
  volume={18},
  number={2},
  pages={41-50},
  doi={10.1109/MNANO.2024.3358714}}

@article{SUN.2020,
title = {Recent progress in SLID bonding in novel 3D-IC technologies},
journal = {J. Alloys Compd.},
volume = {818},
pages = {152825},
year = {2020},
issn = {0925-8388},
doi = {https://doi.org/10.1016/j.j     allcom.2019.152825}          ,
url = {https://www.sciencedirect.com/science/article/pii/S092583881934071X},
author = {Lei Sun and Ming-he Chen and Liang Zhang and Peng He and Lan-sheng Xie}
}

@Article{Wang.2023,
AUTHOR = {Wang, Haoyu and Ma, Jianshe and Yang, Yide and Gong, Mali and Wang, Qinheng},
TITLE = {A Review of System-in-Package Technologies: Application and Reliability of Advanced Packaging},
JOURNAL = {Micromachines},
VOLUME = {14},
YEAR = {2023},
NUMBER = {6},
ARTICLE-NUMBER = {1149},
URL = {https://www.mdpi.com/2072-666X/14/6/1149},
PubMedID = {37374734},
ISSN = {2072-666X},
DOI = {10.3390/mi14061149}
}

@article{Malinowski.2023,
  title = {How to Wire a $1000$-Qubit Trapped-Ion Quantum Computer},
  author = {Malinowski, M. and Allcock, D.T.C. and Ballance, C.J.},
  journal = {PRX Quantum},
  volume = {4},
  issue = {4},
  pages = {040313},
  numpages = {21},
  year = {2023},
  month = {Oct},
  publisher = {American Physical Society},
  doi = {10.1103/PRXQuantum.4.040313},
  url = {https://link.aps.org/doi/10.1103/PRXQuantum.4.040313}            
}

@article{Hu.2022,
 author = {Hu, Tie and Feng, Xing and Yang, Zhenyu and Zhao, Ming},
 year = {2022},
 title = {Design of scalable metalens array for optical addressing},
 pages = {32},
 volume = {15},
 number = {1},
 journal = {Front. Optoelectron.},
 doi = {10.1007/s12200-022-00035-2}
}

@article{Bautista-Salvador.2019,
doi = {10.1088/1367-2630/ab0e46},
url = {https://doi.org/10.1088/1367-2630/ab0e46}                      ,
year = {2019},
month = {apr},
publisher = {IOP Publishing},
volume = {21},
number = {4},
pages = {043011},
author = {Bautista-Salvador, A and Zarantonello, G and Hahn, H and Preciado-Grijalva, A and Morgner, J and Wahnschaffe, M and Ospelkaus, C},
title = {Multilayer ion trap technology for scalable quantum computing and quantum simulation},
journal = {New J. Phys.}
}

@article{Dietl.2025, 
author = {Dietl, Matthias and Valentini, Marco and Anmasser, Fabian and Zesar, Alexander and Auchter, Silke and van Mourik, Martin and Monz, Thomas and Blatt, Rainer and Rössler, Clemens and Schindler, Philipp},
title = {Test and Characterization of Multilayer Ion Traps on Fused Silica},
journal = {Advanced Quantum Technologies},
volume = {8},
number = {11},
doi = {10.1002/qute.202500412},
year = {2025}
}

@INPROCEEDINGS{Coudrain.2019,
  author={Coudrain, Perceval and Charbonnier, J. and Garnier, A. and Vivet, P. and Vélard, Rémi and Vinci, A. and Ponthenier, F. and Farcy, A. and Segaud, R. and Chausse, P. and Arnaud, L. and Lattard, D. and Guthmuller, E. and Romano, G. and Gueugnot, A. and Berger, F. and Beltritti, J. and Mourier, T. and Gottardi, M. and Minoret, S. et al.},
 booktitle={2019 IEEE 69th ECTC},   
 title={Active Interposer Technology for Chiplet-Based Advanced 3D System Architectures}, 
  year={2019},
  volume={},
  number={},
  pages={569-578},
  doi={10.1109/ECTC.2019.00092}}

@article{Schmid.2021,
author = {Michael Schmid and Florian Sterl and Simon Thiele and Alois Herkommer and Harald Giessen},
journal = {Opt. Lett.},
number = {10},
pages = {2485--2488},
publisher = {Optica Publishing Group},
title = {3D printed hybrid refractive/diffractive achromat and apochromat for the visible wavelength range},
volume = {46},
month = {May},
year = {2021},
url = {https://opg.optica.org/ol/abstract.cfm?URI=ol-46-10-2485},
doi = {10.1364/OL.423196},
}

@article{Proctor.2025,
 author = {Proctor, Timothy and Young, Kevin and Baczewski, Andrew D. and Blume-Kohout, Robin},
 year = {2025},
 title = {Benchmarking quantum computers},
 pages = {105--118},
 volume = {7},
 number = {2},
 issn = {2522-5820},
 journal = {Nature Rev. Phys.},
 doi = {10.1038/s42254-024-00796-z}
}

@article{Teller.2021,
  title = {Heating of a Trapped Ion Induced by Dielectric Materials},
  author = {Teller, Markus and Fioretto, Dario A. and Holz, Philip C. and Schindler, Philipp and Messerer, Viktor and Sch\"uppert, Klemens and Zou, Yueyang and Blatt, Rainer and Chiaverini, John and Sage, Jeremy and Northup, Tracy E.},
  journal = {Phys. Rev. Lett.},
  volume = {126},
  issue = {23},
  pages = {230505},
  numpages = {6},
  year = {2021},
  month = {Jun},
  publisher = {American Physical Society},
  doi = {10.1103/PhysRevLett.126.230505},
  url = {https://link.aps.org/doi/10.1103/PhysRevLett.126.230505}             
}

@article{Clark.1968,
title = {Low temperature thermal expansion of some metallic alloys},
journal = {Cryogenics},
volume = {8},
number = {5},
pages = {282-289},
year = {1968},
issn = {0011-2275},
doi = {https://doi.org/10.1016/S0011-2275(68)80003-7}            ,
url = {https://www.sciencedirect.com/science/article/pii/S0011227568800037},
author = {A.F. Clark}
}

@misc{knollmann.2025,
      title={Collection of fluorescence from an ion using trap-integrated photonics}, 
      author={Felix W. Knollmann and Sabrina M. Corsetti and Ethan R. Clements and Reuel Swint and Aaron D. Leu and May E. Kim and Patrick T. Callahan and Dave Kharas and Thomas Mahony and Cheryl Sorace-Agaskar and Robert McConnell and Colin D. Bruzewicz and Isaac L. Chuang and Jelena Notaros and John Chiaverini},
      year={2025},
      eprint={2505.01412},
      archivePrefix={arXiv},
}

@Article{Charaev.2023,
author={Charaev, I.
and Bandurin, D. A.
and Bollinger, A. T.
and Phinney, I. Y.
and Drozdov, I.
and Colangelo, M.
and Butters, B. A.
and Taniguchi, T.
and Watanabe, K.
and He, X.
and Medeiros, O.
and Bo{\v{z}}ovi{\'{c}}, I.
and Jarillo-Herrero, P.
and Berggren, K. K.},
title={Single-photon detection using high-temperature superconductors},
journal={Nat. Nanotechnol.},
year={2023},
month={Apr},
day={01},
volume={18},
number={4},
pages={343-349},
issn={1748-3395},
doi={10.1038/s41565-023-01325-2},
url={https://doi.org/10.1038/s41565-023-01325-2}           
}

@Article{Charaev.2024,
author={Charaev, Ilya
and Batson, Emma K.
and Cherednichenko, Sergey
and Reidy, Kate
and Drakinskiy, Vladimir
and Yu, Yang
and Lara-Avila, Samuel
and Thomsen, Joachim D.
and Colangelo, Marco
and Incalza, Francesca
and Ilin, Konstantin
and Schilling, Andreas
and Berggren, Karl K.},
title={Single-photon detection using large-scale high-temperature MgB2 sensors at 20 K},
journal={Nat. Commun.},
year={2024},
month={May},
day={10},
volume={15},
number={1},
pages={3973},
issn={2041-1723},
doi={10.1038/s41467-024-47353-x},
url={https://doi.org/10.1038/s41467-024-47353-x}        
}

@article{Guise.2015,
author = {Guise, Nicholas D. and Fallek, Spencer D. and Stevens, Kelly E. and Brown, K. R. and Volin, Curtis and Harter, Alexa W. and Amini, Jason M. and Higashi, Robert E. and Lu, Son Thai and Chanhvongsak, Helen M. and Nguyen, Thi A. and Marcus, Matthew S. and Ohnstein, Thomas R. and Youngner, Daniel W.},
title = {Ball-grid array architecture for microfabricated ion traps},
journal = {J. Appl. Phys.},
volume = {117},
number = {17},
pages = {174901},
year = {2015},
month = {05},
issn = {0021-8979},
doi = {10.1063/1.4917385},
url = {https://doi.org/10.1063/1.4917385}       ,
}

@Article{Stick.2006,
author={Stick, D.
and Hensinger, W. K.
and Olmschenk, S.
and Madsen, M. J.
and Schwab, K.
and Monroe, C.},
title={Ion trap in a semiconductor chip},
journal={Nature Physics},
year={2006},
month={Jan},
day={01},
volume={2},
number={1},
pages={36-39},
issn={1745-2481},
doi={10.1038/nphys171},
url={https://doi.org/10.1038/nphys171}        
}

@article{Chiaverini.2005,
author = {Chiaverini, J. and Blakestad, R. B. and Britton, J. and Jost, J. D. and Langer, C. and Leibfried, D. and Ozeri, R. and Wineland, D. J.},
title = {Surface-electrode architecture for ion-trap quantum information processing},
year = {2005},
issue_date = {September 2005},
publisher = {Rinton Press, Incorporated},
address = {Paramus, NJ},
volume = {5},
number = {6},
issn = {1533-7146},
journal = {Quantum Info. Comput.},
month = sep,
pages = {419–439},
numpages = {21},
doi = {10.26421/QIC5.6-1}
}

@Article{Akhtar.2023,
author={Akhtar, M.
and Bonus, F.
and Lebrun-Gallagher, F. R.
and Johnson, N. I.
and Siegele-Brown, M.
and Hong, S.
and Hile, S. J.
and Kulmiya, S. A.
and Weidt, S.
and Hensinger, W. K.},
title={A high-fidelity quantum matter-link between ion-trap microchip modules},
journal={Nat. Commun.},
year={2023},
month={Feb},
day={08},
volume={14},
number={1},
pages={531},
issn={2041-1723},
doi={10.1038/s41467-022-35285-3},
url={https://doi.org/10.1038/s41467-022-35285-3         }
}

@article{Lekitsch.2017,
author = {Bjoern Lekitsch  and Sebastian Weidt  and Austin G. Fowler  and Klaus Mølmer  and Simon J. Devitt  and Christof Wunderlich  and Winfried K. Hensinger },
title = {Blueprint for a microwave trapped ion quantum computer},
journal = {Sci. Adv.},
volume = {3},
number = {2},
pages = {e1601540},
year = {2017},
doi = {10.1126/sciadv.1601540},
URL = {https://www.science.org/doi/abs/10.1126/sciadv.1601540}}

@Article{Romaszko2020,
author={Romaszko, Zak David
and Hong, Seokjun
and Siegele, Martin
and Puddy, Reuben Kahan
and Lebrun-Gallagher, Foni Rapha{\"e}l
and Weidt, Sebastian
and Hensinger, Winfried Karl},
title={Engineering of microfabricated ion traps and integration of advanced on-chip features},
journal={Nature Rev. Phys.},
year={2020},
pages = {278--295},
month={Jun},
day={01},
volume={2},
doi = {10.1038/s42254-020-0182-8}
}

@article{Auchter.2022,
 author = {Auchter, S. and Axline, C. and Decaroli, C. and Valentini, M. and Purwin, L. and Oswald, R. and Matt, R. and Aschauer, E. and Colombe, Y. and Holz, P. and Monz, T. and Blatt, R. and Schindler, P. and R{\"o}ssler, C. and Home, J.},
 year = {2022},
 title = {Industrially microfabricated ion trap with 1 eV trap depth},
 pages = {035015},
 volume = {7},
 number = {3},
 journal = {JQST},
 doi = {10.1088/2058-9565/ac7072 }
}

@article{Blain.2021,
doi = {10.1088/2058-9565/ac01bb },
url = {https://dx.doi.org/10.1088/2058-9565/ac01bb     }              ,
year = {2021},
month = {jun},
publisher = {IOP Publishing},
volume = {6},
number = {3},
pages = {034011},
author = {Blain, M G and Haltli, R and Maunz, P and Nordquist, C D and Revelle, M and Stick, D},
title = {Hybrid MEMS-CMOS ion traps for NISQ computing},
journal = {JQST}
}

@article{Blakestad.2009,
 author = {Blakestad, R. B. and Ospelkaus, C. and VanDevender, A. P. and Amini, J. M. and Britton, J. and Leibfried, D. and Wineland, D. J.},
 year = {2009},
 title = {High-fidelity transport of trapped-ion qubits through an X-junction trap array},
 pages = {153002},
 volume = {102},
 number = {15},
 journal = {Phys. Rev. Lett.},
 doi = {10.1103/PhysRevLett.102.153002}
}

@article{Brown.2021,
    author = {Brown, Kenneth R. and Chiaverini, John and Sage, Jeremy M. and Häffner, Hartmut},
    title = {Materials challenges for trapped-ion quantum computers},
    journal = {Nature Rev. Mat.},
    year = {2021},
    pages = {892-905},
    volume = {6},
    number = {10},
    url = {https://doi.org/10.1038/s41578-021-00292-1}              ,
    doi = {10.1038/s41578-021-00292-1}
}

@article{Bruzewicz.2015,
 author = {Bruzewicz, C. D. and Sage, J. M. and Chiaverini, J.},
 year = {2015},
 title = {Measurement of ion motional heating rates over a range of trap frequencies and temperatures},
 volume = {91},
 number = {4},
 issn = {2469-9926},
 journal = {Phys. Rev. A},
 doi = {10.1103/PhysRevA.91.041402}
}

@article{Cho.2015,
 author = {Cho, Dong-Il ``Dan'' and Hong, Seokjun and Lee, Minjae and Kim, Taehyun},
 year = {2015},
 title = {A review of silicon microfabricated ion traps for quantum information processing},
 volume = {3},
 number = {1},
 journal = {MNSL},
 doi = {10.1186/s40486-015-0013-3}
}

@article{Daniilidis.2014,
 author = {Daniilidis, N. and Gerber, S. and Bolloten, G. and Ramm, M. and Ransford, A. and Ulin-Avila, E. and Talukdar, I. and H{\"a}ffner, H.},
 year = {2014},
 title = {Surface noise analysis using a single-ion sensor},
 volume = {89},
 number = {24},
 issn = {1098-0121},
 journal = {Phys. Rev. B},
 doi = {10.1103/PhysRevB.89.245435}
}

@inproceedings{Graef.2021,
 author = {Graef, Mart},
 title = {More Than Moore White Paper},
 pages = {1--47},
 publisher = {IEEE},
 isbn = {978-1-6654-8638-5},
 booktitle = {2021 IEEE IRDS Outbriefs},
 year = {2021},
 doi = {10.1109/IRDS54852.2021.00013}
}

@article{Hampel.2023,
 author = {Hampel, Benedikt and Slichter, Daniel H. and Leibfried, Dietrich and Mirin, Richard P. and Nam, Sae Woo and Verma, Varun B.},
 year = {2023},
 title = {Trap-Integrated Superconducting Nanowire Single-Photon Detectors with Improved RF Tolerance for Trapped-Ion Qubit State Readout},
 volume = {122},
 number = {17},
 issn = {0003-6951},
 journal = {Appl. Phys. Lett.},
 doi = {10.1063/5.0145077}
}

@article{Hite.2012,
 author = {Hite, D. A. and Colombe, Y. and Wilson, A. C. and Brown, K. R. and Warring, U. and J{\"o}rdens, R. and Jost, J. D. and McKay, K. S. and Pappas, D. P. and Leibfried, D. and Wineland, D. J.},
 year = {2012},
 title = {100-fold reduction of electric-field noise in an ion trap cleaned with in situ argon-ion-beam bombardment},
 pages = {103001},
 volume = {109},
 number = {10},
 journal = {Phys. Rev. Lett.},
 doi = {10.1103/PhysRevLett.109.103001}
}

@article{Hogle.2023,
 author = {Hogle, C. W. and Dominguez, D. and Dong, M. and Leenheer, A. and McGuinness, H. J. and Ruzic, B. P. and Eichenfield, M. and Stick, D.},
 year = {2023},
 title = {High-fidelity trapped-ion qubit operations with scalable photonic modulators},
 volume = {9},
 number = {1},
 journal = {npj Quantum Information},
 doi = {10.1038/s41534-023-00737-1}
}

@article{Holz.2020,
 author = {Holz, Philip C. and Auchter, Silke and Stocker, Gerald and Valentini, Marco and Lakhmanskiy, Kirill and R{\"o}ssler, Clemens and Stampfer, Paul and Sgouridis, Sokratis and Aschauer, Elmar and Colombe, Yves and Blatt, Rainer},
 year = {2020},
 title = {2D Linear Trap Array for Quantum Information Processing},
 volume = {3},
 number = {11},
 issn = {2511-9044},
 journal = {Advanced Quantum Technologies},
 doi = {10.1002/qute.202000031}
}

@article{James.1998,
 author = {James, D.F.V.},
 year = {1998},
 title = {Quantum dynamics of cold trapped ions with application to quantum computation},
 url = {https://link.springer.com/article/10.1007/s003400050373}         ,
 pages = {181--190},
 volume = {66},
 number = {2},
 issn = {1432-0649},
 journal = {Appl. Phys. B.},
 doi = {10.1007/s003400050373}
}

@article{Kazuhiro.2025,
  title = {Simulating Floquet scrambling circuits on trapped-ion quantum computers},
  author = {Seki, Kazuhiro and Kikuchi, Yuta and Hayata, Tomoya and Yunoki, Seiji},
  journal = {Phys. Rev. Res.},
  volume = {7},
  issue = {2},
  pages = {023032},
  numpages = {25},
  year = {2025},
  month = {Apr},
  publisher = {American Physical Society},
  doi = {10.1103/PhysRevResearch.7.023032},
  url = {https://link.aps.org/doi/10.1103/PhysRevResearch.7.023032}           
}

@article{Khromova.2012,
  title = {Designer Spin Pseudomolecule Implemented with Trapped Ions in a Magnetic Gradient},
  author = {Khromova, A. and Piltz, Ch. and Scharfenberger, B. and Gloger, T. F. and Johanning, M. and Var\'on, A. F. and Wunderlich, Ch.},
  journal = {Phys. Rev. Lett.},
  volume = {108},
  issue = {22},
  pages = {220502},
  numpages = {5},
  year = {2012},
  month = {Jun},
  publisher = {American Physical Society},
  doi = {10.1103/PhysRevLett.108.220502},
  url = {https://link.aps.org/doi/10.1103/PhysRevLett.108.220502}   
}

@article{Kielpinski.2002,
 author = {Kielpinski, D. and Monroe, C. and Wineland, D. J.},
 year = {2002},
 title = {Architecture for a large-scale ion-trap quantum computer},
 pages = {709--711},
 volume = {417},
 number = {6890},
 issn = {1476-4687},
 journal = {Nature},
 doi = {10.1038/nature00784}
}

@article{Knollmann.2024,
 author = {Knollmann, F. W. and Clements, E. and Callahan, P. T. and Gehl, M. and Hunker, J. D. and Mahony, T. and McConnell, R. and Swint, R. and Sorace-Agaskar, C. and Chuang, I. L. and Chiaverini, J. and Stick, D.},
 year = {2024},
 title = {Integrated photonic structures for photon-mediated entanglement of trapped ions},
 url = {http://arxiv.org/pdf/2401.06850},
 pages = {230},
 volume = {2},
 number = {4},
 journal = {Optica Quantum},
 doi = {10.1364/OPTICAQ.522128}
}

@misc{Kumar.3132025,
      title={Fast-response low power atomic oven for integration into an ion microchip}, 
      author={Vijay Kumar and Martin Siegele-Brown and Parsa Rahimi and Matthew Aylett and Sebastian Weidt and Winfried Karl Hensinger},
      year={2025},
      eprint={2503.10550},
      archivePrefix={arXiv},
}

@article{Kwon.2024,
 author = {Kwon, Joonhyuk and Setzer, William J. and Gehl, Michael and Karl, Nicholas and {van der Wall}, Jay and Law, Ryan and Blain, Matthew G. and Stick, Daniel and McGuinness, Hayden J.},
 year = {2024},
 title = {Multi-site integrated optical addressing of trapped ions},
 pages = {3709},
 volume = {15},
 number = {1},
 journal = {Nat. Commun.},
 doi = {10.1038/s41467-024-47882-5}
}

@inproceedings{Mehta.2023,
 author = {Mehta, Karan and Ricci-Vasquez, Alfredo and Mordini, Carmelo and Beck, Gillenhaal and Malinowski, Maciej and Stadler, Martin and Zhang, Chi and Kienzler, Daniel and Home, Jonathan},
 title = {Ion trap quantum computing using integrated photonics},
 url = {https://www.spiedigitallibrary.org/conference-proceedings-of-spie/12424/2655382/Ion-trap-quantum-computing-using-integrated-photonics/10.1117/12.2655382.full}                ,
 pages = {33},
 publisher = {SPIE},
 isbn = {9781510659537},
 series = {Proceedings of SPIE},
 editor = {Garc{\'i}a-Blanco, Sonia M. and Cheben, Pavel},
 booktitle = {Integrated Optics: Devices, Materials, and Technologies XXVII},
 year = {2023},
 address = {Bellingham, Washington, USA},
 doi = {10.1117/12.2655382}
}

@article{Mehta.2016,
 author = {Mehta, Karan K. and Bruzewicz, Colin D. and McConnell, Robert and Ram, Rajeev J. and Sage, Jeremy M. and Chiaverini, John},
 year = {2016},
 title = {Integrated optical addressing of an ion qubit},
 pages = {1066--1070},
 volume = {11},
 number = {12},
 journal = {Nat. Nanotechnol.},
 doi = {10.1038/nnano.2016.139}
}

@article{Mehta.2020,
 author = {Mehta, Karan K. and Zhang, Chi and Malinowski, Maciej and Nguyen, Thanh-Long and Stadler, Martin and Home, Jonathan P.},
 year = {2020},
 title = {Integrated optical multi-ion quantum logic},
 pages = {533--537},
 volume = {586},
 number = {7830},
 issn = {1476-4687},
 journal = {Nature},
 doi = {10.1038/s41586-020-2823-6}
}

@article{Monroe.2014,
  title = {Large-scale modular quantum-computer architecture with atomic memory and photonic interconnects},
  author = {Monroe, C. and Raussendorf, R. and Ruthven, A. and Brown, K. R. and Maunz, P. and Duan, L.-M. and Kim, J.},
  journal = {Phys. Rev. A},
  volume = {89},
  issue = {2},
  pages = {022317},
  numpages = {16},
  year = {2014},
  month = {Feb},
  publisher = {American Physical Society},
  doi = {10.1103/PhysRevA.89.022317},
  url = {https://link.aps.org/doi/10.1103/PhysRevA.89.022317}       
}

@article{Moore.2006,
 author = {Moore, Gordon E.},
 year = {2006},
 title = {Cramming more components onto integrated circuits, Reprinted from Electronics, volume 38, number 8, April 19, 1965, pp.114 ff},
 pages = {33--35},
 volume = {11},
 number = {3},
 issn = {1098-4232},
 journal = {IEEE SSC-M},
 doi = {10.1109/N-SSC.2006.4785860}
}

@article{Moore.2006b,
 author = {Moore, Gordon E.},
 year = {2006},
 title = {Progress in digital integrated electronics [Technical literaiture, Copyright 1975 IEEE. Reprinted, with permission. Technical Digest. International Electron Devices Meeting, IEEE, 1975, pp. 11-13.]},
 pages = {36--37},
 volume = {11},
 number = {3},
 issn = {1098-4232},
 journal = {IEEE SSC-M},
 doi = {10.1109/N-SSC.2006.4804410}
}

@article{Moses.2023,
 author = {Moses, S. A. and Baldwin, C. H. and Allman, M. S. and Ancona, R. and Ascarrunz, L. and Barnes, C. and Bartolotta, J. and Bjork, B. and Blanchard, P. and Bohn, M. and Bohnet, J. G. and Brown, N. C. and Burdick, N. Q. and Burton, W. C. and Campbell, S. L. and Campora, J. P. and Carron, C. and Chambers, J. and Chan, J. W. and Chen, Y. H. et al.},
 year = {2023},
 title = {A Race-Track Trapped-Ion Quantum Processor},
 volume = {13},
 number = {4},
 journal = {Phys. Rev. X},
 doi = {10.1103/PhysRevX.13.041052}
}

@article{Niffenegger.2020,
 author = {Niffenegger, R. J. and Stuart, J. and Sorace-Agaskar, C. and Kharas, D. and Bramhavar, S. and Bruzewicz, C. D. and Loh, W. and Maxson, R. T. and McConnell, R. and Reens, D. and West, G. N. and Sage, J. M. and Chiaverini, J.},
 year = {2020},
 title = {Integrated multi-wavelength control of an ion qubit},
 pages = {538--542},
 volume = {586},
 number = {7830},
 issn = {1476-4687},
 journal = {Nature},
 doi = {10.1038/s41586-020-2811-x}
}

@article{OHalloran.2023,
 author = {O'Halloran, Se{\'a}n and Pandit, Abhay and Heise, Andreas and Kellett, Andrew},
 year = {2023},
 title = {Two-Photon Polymerization: Fundamentals, Materials, and Chemical Modification Strategies},
 pages = {e2204072},
 volume = {10},
 number = {7},
 journal = {Adv. Sci. (Weinh).},
 doi = {10.1002/advs.202204072}
}

@article{Ollitrault.2024.04.24,
    author =  {Ollitrault, Pauline J. and Loipersberger, Matthias and Parrish, Robert M. and Erhard, Alexander and Maier, Christine and Sommer, Christian and Ulmanis, Juris and Monz, Thomas and Gogolin, Christian and Tautermann, Christofer S. and Anselmetti, Gian-Luca R. and Degroote, Matthias and Moll, Nikolaj and Santagati, Raffaele and Streif, Michael
},
    title = {Estimation of Electrostatic Interaction Energies on a Trapped-Ion Quantum Computer},
    journal = {ACS Cent. Sci.},
    volume = {10},
    number = {4},
    pages = {882-889},
    year = {2024},
    doi = {10.1021/acscentsci.4c00058},
    url = {https://doi.org/10.1021/acscentsci.4c00058}   
}

@article{Pino.2021,
 author = {Pino, J. M. and Dreiling, J. M. and Figgatt, C. and Gaebler, J. P. and Moses, S. A. and Allman, M. S. and Baldwin, C. H. and Foss-Feig, M. and Hayes, D. and Mayer, K. and Ryan-anderson, C. and Neyenhuis, B.},
 year = {2021},
 title = {Demonstration of the trapped-ion quantum CCD computer architecture},
 url = {https://www.nature.com/articles/s41586-021-03318-4#citeas},
 pages = {209--213},
 volume = {592},
 number = {7853},
 issn = {1476-4687},
 journal = {Nature},
 doi = {10.1038/s41586-021-03318-4}
}

@article{Postler.2022,
 author = {Postler, Lukas and Heu{\ss}en, Sascha and Pogorelov, Ivan and Rispler, Manuel and Feldker, Thomas and Meth, Michael and Marciniak, Christian D. and Stricker, Roman and Ringbauer, Martin and Blatt, Rainer and Schindler, Philipp and M{\"u}ller, Markus and Monz, Thomas},
 year = {2022},
 title = {Demonstration of fault-tolerant universal quantum gate operations},
 pages = {675--680},
 volume = {605},
 number = {7911},
 issn = {1476-4687},
 journal = {Nature},
 doi = {10.1038/s41586-022-04721-1}
}

@book{Ramm.2012,
 author = {Ramm, Peter and Lu, James Jian--Qiang and Taklo, Maaike M. V.},
 year = {2012},
 title = {Handbook of wafer bonding},
 address = {Weinheim},
 publisher = {Wiley-VCH-Verl.},
 doi = {10.1002/9783527644223}
}

@article{Reens.2022,
 author = {Reens, David and Collins, Michael and Ciampi, Joseph and Kharas, Dave and Aull, Brian F. and Donlon, Kevan and Bruzewicz, Colin D. and Felton, Bradley and Stuart, Jules and Niffenegger, Robert J. and Rich, Philip and Braje, Danielle and Ryu, Kevin K. and Chiaverini, John and McConnell, Robert},
 year = {2022},
 title = {High-Fidelity Ion State Detection Using Trap-Integrated Avalanche Photodiodes},
 pages = {100502},
 volume = {129},
 number = {10},
 journal = {Phys. Rev. Lett.},
 doi = {10.1103/PhysRevLett.129.100502}
}

@article{Ryananderson.19.09.2024,
author = {C. Ryan-Anderson and N. C. Brown  and C. H. Baldwin  and J. M. Dreiling  and C. Foltz  and J. P. Gaebler  and T. M. Gatterman  and N. Hewitt  and C. Holliman  and C. V. Horst  and J. Johansen  and D. Lucchetti  and T. Mengle  and M. Matheny  and Y. Matsuoka  and K. Mayer  and M. Mills  and S. A. Moses  and B. Neyenhuis  and J. Pino  et. al},
title = {High-fidelity teleportation of a logical qubit using transversal gates and lattice surgery},
journal = {Science},
volume = {385},
number = {6715},
pages = {1327-1331},
year = {2024},
doi = {10.1126/science.adp6016},
URL = {https://www.science.org/doi/abs/10.1126/science.adp6016},
}

@article{Setzer.2021,
author = {Setzer, W. J. and Ivory, M. and Slobodyan, O. and Van Der Wall, J. W. and Parazzoli, L. P. and Stick, D. and Gehl, M. and Blain, M. G. and Kay, R. R. and McGuinness, H. J.},
    title = {Fluorescence detection of a trapped ion with a monolithically integrated single-photon-counting avalanche diode},
    journal = {Appl. Phys. Lett.},
    volume = {119},
    number = {15},
    pages = {154002},
    year = {2021},
    month = {10},
    issn = {0003-6951},
    doi = {10.1063/5.0055999},
    url = {https://doi.org/10.1063/5.0055999},
}

@inproceedings{Sieberer.2021,
 author = {Sieberer, Michael and Sandner, Christoph and Hadley, Peter},
 title = {A Cryogenic High-Voltage Amplifier for Ion Traps},
 pages = {1--4},
 booktitle = {SMACD / PRIME 2021},
 year = {2021},
 isbn = {978-3-8007-5588-2},
 url = {https://ieeexplore.ieee.org/document/9548003}
}

@inproceedings{Simeth.1282023232023,
author = {Sebastian J. Simeth and Alexander M{\"u}ller and Jan M{\"u}ller and Bj{\"o}rn Lekitsch and Martin Reininghaus and Ferdinand Schmidt-Kaler},
title = {{Selective laser-induced etching for 3D ion traps}},
volume = {12409},
booktitle = {Laser-based Micro- and Nanoprocessing XVII},
editor = {Rainer Kling and Akira Watanabe and Wilhelm Pfleging},
organization = {International Society for Optics and Photonics},
publisher = {SPIE},
pages = {1240902},
year = {2023},
doi = {10.1117/12.2647189},
URL = {https://doi.org/10.1117/12.2647189}
}

@article{Smith.2025,
  title = {Single-Qubit Gates with Errors at the ${10}^{\ensuremath{-}7}$ Level},
  author = {Smith, M. C. and Leu, A. D. and Miyanishi, K. and Gely, M. F. and Lucas, D. M.},
  journal = {Phys. Rev. Lett.},
  volume = {134},
  issue = {23},
  pages = {230601},
  numpages = {8},
  year = {2025},
  month = {Jun},
  publisher = {American Physical Society},
  doi = {10.1103/42w2-6ccy},
  url = {https://link.aps.org/doi/10.1103/42w2-6ccy}
}

@misc{Sterk.2024,
      title={Multi-junction surface ion trap for quantum computing}, 
      author={J. D. Sterk and M. G. Blain and M. Delaney and R. Haltli and E. Heller and A. L. Holterhoff and T. Jennings and N. Jimenez and A. Kozhanov and Z. Meinelt and E. Ou and J. Van Der Wall and C. Noel and D. Stick},
      year={2024},
      eprint={2403.00208},
      archivePrefix={arXiv},
}

@article{Stuart.2019,
 author = {Stuart, J. and Panock, R. and Bruzewicz, C. D. and Sedlacek, J. A. and McConnell, R. and Chuang, I. L. and Sage, J. M. and Chiaverini, J.},
 year = {2019},
 title = {Chip-Integrated Voltage Sources for Control of Trapped Ions},
 pages = {024010},
 volume = {11},
 number = {2},
 journal = {Phys. Rev. Applied},
 doi = {10.1103/PhysRevApplied.11.024010}
}

@article{Tinkey.2022,
 author = {Tinkey, Holly N. and Clark, Craig R. and Sawyer, Brian C. and Brown, Kenton R.},
 year = {2022},
 title = {Transport-Enabled Entangling Gate for Trapped Ions},
 pages = {050502},
 volume = {128},
 number = {5},
 journal = {Phys. Rev. Lett.},
 doi = {10.1103/PhysRevLett.128.050502}
}

@article{Todaro.2021,
 author = {Todaro, S. L. and Verma, V. B. and McCormick, K. C. and Allcock, D. T. C. and Mirin, R. P. and Wineland, D. J. and Nam, S. W. and Wilson, A. C. and Leibfried, D. and Slichter, D. H.},
 year = {2021},
 title = {State Readout of a Trapped Ion Qubit Using a Trap-Integrated Superconducting Photon Detector},
 pages = {010501},
 volume = {126},
 number = {1},
 journal = {Phys. Rev. Lett.},
 doi = {10.1103/PhysRevLett.126.010501}
}

@article{Valentini.2025,
  title = {Demonstration of Two-Dimensional Connectivity for a Scalable Error-Corrected Ion-Trap Quantum Processor Architecture},
  author = {Valentini, M. and van Mourik, M. W. and Butt, F. and Wahl, J. and Dietl, M. and Pfeifer, M. and Anmasser, F. and Colombe, Y. and R\"ossler, C. and Holz, P. C. and Blatt, R. and Bermudez, A. and M\"uller, M. and Monz, T. and Schindler, P.},
  journal = {Phys. Rev. X},
  volume = {15},
  issue = {4},
  pages = {041023},
  numpages = {32},
  year = {2025},
  month = {Nov},
  publisher = {American Physical Society},
  doi = {10.1103/b9s1-6r44},
  url = {https://link.aps.org/doi/10.1103/b9s1-6r44}  
}

@article{Wang.11.01.2021, 
 author = {Wang, Pengfei and Luan, Chun-Yang and Qiao, Mu and Um, Mark and Zhang, Junhua and Wang, Ye and Yuan, Xiao and Gu, Mile and Zhang, Jingning and Kim, Kihwan},
 year = {2021},
 title = {Single ion qubit with estimated coherence time exceeding one hour},
 pages = {233},
 volume = {12},
 number = {1},
 journal = {Nat. Commun.},
 doi = {10.1038/s41467-020-20330-w},
 url = {https://doi.org/10.1038/s41467-020-20330-w}
}

@article{Zhu.2022,
doi = {10.1088/2058-9565/ac91ef},
url = {https://dx.doi.org/10.1088/2058-9565/ac91ef} ,
year = {2022},
month = {nov},
publisher = {IOP Publishing},
volume = {8},
number = {1},
pages = {015007},
author = {Zhu, Yingyue and Zhang, Zewen and Sundar, Bhuvanesh and Green, Alaina M and Huerta Alderete, C and Nguyen, Nhung H and Hazzard, Kaden R A and Linke, Norbert M},
title = {Multi-round QAOA and advanced mixers on a trapped-ion quantum computer},
journal = {JQST}
}

@article{Zhu.2023,
doi = {10.1038/s41598-023-44151-1},
url = {https://doi.org/10.1038/s41598-023-44151-1} ,
year = {2023},
month = {},
publisher = {},
volume = {13},
number = {1},
pages = {18511},
author = {Zhu, Daiwei and Shen, Weiwei and Giani, Annarita and Ray-Majumder, Saikat and Neculaes, Bogdan and Johri, Sonika},
title = {Copula-based risk aggregation with trapped ion quantum computers},
journal = {Sci. Rep.}
}

\end{document}